\documentclass[aps,pre,preprint]{revtex4}

\usepackage{latexsym}
\usepackage{amsmath}
\usepackage{amssymb}
\usepackage{mathtools}
\usepackage{bm}

\usepackage{graphicx}
\usepackage{epsfig}
\usepackage{epstopdf}
\usepackage[caption=false]{subfig}

\usepackage{placeins}
\usepackage{xcolor}

\providecommand{\repeatdots}[1]{
	\hbox{.} 
	\ifnum1<\numexpr#1\relax
		\sdots{#1-1}
	\fi
}
\providecommand{\sdots}[1]{%
	\vbox{
		\baselineskip 0ex \lineskip 0.2ex
		\repeatdots{#1}
	}%
}
\providecommand{\symmetric}[3]{%
	\setbox0\hbox{\ensuremath{#1}}
	\vbox{
		\baselineskip 0ex \lineskip 0.25ex
		\hbox to \wd0{%
			\hss\hspace{0.2\ht0}\sdots{#2}\hspace{-0.15ex}\sdots{#3}\hss%
		}
		\box0
	}
	\vphantom{#1}
}

\makeatletter
\renewcommand*\env@matrix[1][\arraystretch]{%
	\edef\arraystretch{#1}%
	\hskip -\arraycolsep
	\let\@ifnextchar\new@ifnextchar
	\array{*\c@MaxMatrixCols c}}
\makeatother

\renewcommand{\vec}[1]{\boldsymbol{#1}}
\providecommand{\mat}[1]{\boldsymbol{#1}}

\providecommand{\diff}{\mathrm{d}}

\providecommand{\abs}[1]{\lvert#1\rvert}

\providecommand{\eqref}[1]{(\ref{#1})}
\providecommand{\vanish}[1]{}

\providecommand{\cs}{\ensuremath{\mathrm{c}}}
\providecommand{\sn}{\ensuremath{\mathrm{s}}}
\providecommand{\so}[1]{\ensuremath{\mathrm{SO({#1})}}}
\providecommand{\su}[1]{\ensuremath{\mathrm{SU({#1})}}}

\begin{document}

\title{Basis functions on the grain boundary space: Theory}

\author{J. K. Mason}
\email{jkmason@ucdavis.edu}
\affiliation{University of California, Davis, CA 95616 United States}
\author{S. Patala}
\email{spatala@ncsu.edu}
\affiliation{North Carolina State University, Raleigh, NC 27695 United States}

\begin{abstract}
With the increasing availability of experimental and computational data concerning the properties and distribution of grain boundaries in polycrystalline materials, there is a corresponding need to efficiently and systematically express functions on the grain boundary space. A grain boundary can be described by the rotations applied to two grains on either side of a fixed boundary plane, suggesting that the grain boundary space is related to the space of rotations. This observation is used to construct an orthornormal function basis, allowing effectively arbitrary functions on the grain boundary space to be written as linear combinations of the basis functions. Moreover, a procedure is developed to construct a smaller set of basis functions consistent with the crystallographic point group symmetries, grain exchange symmetry, and the null boundary singularity. Functions with the corresponding symmetries can be efficiently expressed as linear combinations of the symmetrized basis functions. An example is provided that shows the efficacy of the symmetrization procedure.
\end{abstract}


\maketitle

\section{Introduction}
\label{sec:introduction}

Over the past few decades, materials scientists have increasingly come to the realization that the distribution and connectivity of different grain boundary (GB) types contributes to the mechanical and functional properties of polycrystalline materials \cite{shimada2002optimization, hansen2004hall, lu2004ultrahigh, meyers2006mechanical, randle2010grain, bagri2011thermal, fang2011revealing}. Despite the role of GB structure in transport and failure mechanisms having been investigated for more than half a century \cite{forsyth1946grain, smoluchowski1952theory, haynes1955grain, hirth1972influence, hunderi1973influence, chadwick1976grain, gleiter1981interaction, gleiter1982structure, dimos1990superconducting, sutton1995interfaces, gottstein2009grain}, few robust GB structure-property relationships are yet known; this is at least partly due to the inherent complexity associated with the five-dimensional configuration space in which they reside \cite{morawiec2003orientations, patala2012improved, patala2013symmetries}. The absence of such relationships remains one of the most significant obstacles to developing true bottom-up models for the behavior of polycrystalline materials \cite{panchal2013key}. 

Recent advances in both experimental and computational techniques have facilitated the construction of large databases of GB properties \cite{olmsted2009survey, olmsted2009survey2, holm2010comparing, rohrer2011grain, homer2014trends} in the five-parameter space. These five macroscopic degrees of freedom refer to the misorientation (three parameters) and the boundary-plane inclination (two parameters) of the GB. With the advent of modern high-throughput algorithms \cite{banadaki2018efficient,homer2019high} and sophisticated experimental techniques \cite{2012poulsen,seita2016high}, the community is now at the point where the development of appropriate statistical techniques is critical to the analysis of the vast amounts of data being generated and to the building of predictive models essential to advance the field of GB science and engineering.

The purpose of this article is to introduce a standard approach to representing and manipulating functions on the five-parameter space of grain boundaries, and which involves expanding them over a set of orthonormal basis functions. Basis functions have been proven to be extremely useful in analyzing data in complex domains \cite{walczak1996radial,scholkopf1997comparing,ramamoorthi2001efficient}. A simple case is when the data is on the real line $\mathbb{R}^1$ and is $2 \pi$ periodic (i.e.\ the domain is the unit circle); then the Fourier series or the set of functions $\{e^{imx} : m \in \mathbb{N}\}$ constitutes the natural set of basis functions. Similarly, if the data is on the unit sphere, the spherical harmonics $Y_l^m (\theta, \phi)$ constitute the natural set of basis functions.

The concept of orthonormal basis functions is also central to the field of texture analysis, which is concerned with the measurement and analysis of the probability distribution of crystal orientations in a polycrystalline material. This conventionally involves expanding the probability distribution over a set of basis functions with desirable properties and manipulating the coefficients of the expansion. The basis functions can be constructed to be orthonormal with respect to an inner product, to obey the crystal and sample point group symmetries, and to undergo well-defined transformations when the sample is rotated in the laboratory frame. This does not uniquely define the set of basis functions though, and more than one reasonable choice is available in the literature \cite{1982bunge,2009mason_b}.

We propose that the main criterion to select a set of orthonormal basis functions when crystallography is involved be the convenience of expressing the effect of a rotation of the sample in the laboratory frame as a transformation of the expansion coefficients. The motivation for this is that the action of a rotation on the expansion coefficients allows the basis functions to be symmetrized to obey the crystal and sample point group symmetries. Performing the expansion over the symmetrized basis functions instead of the initial ones allows a texture to be expressed by fewer coefficients, reduces the uncertainty in textures measured by diffraction experiments, and simplifies the graphical representation of texture information \cite{2011patala}. Since the main practical difficulty with this procedure is the complexity of the mathematical expressions involved in defining the symmetrized basis functions, a set of basis functions with a simple transformation rule is greatly desired.

This article applies these ideas to probability distributions of grain boundaries. This has not been done before, to the best of our knowledge. One reason for this is the historical difficulty of experimental measurements of the distribution of grain boundary types in a microstructure. The increasing accessibility of three-dimensional X-ray diffraction microscopy \cite{2012poulsen} and three-dimensional electron backscatter diffraction \cite{2007rollett} promises to make this data more available in the near future. A second likely reason is the complexity of the mathematical expressions involved. A texture can be considered as a probability distribution on \so{3} (the group of rotations in three dimensions), an object that is well-studied and appears in a variety of contexts. A probability distribution of grain boundaries is instead a probability distribution on $\so{3} \times \so{3} / \so{2}$ considered as a topological space. This object is not often discussed in the scientific literature, meaning that the construction of functions on this space does not benefit from the prior work of other researchers.

Section \ref{sec:fun_of_rotations} introduces two functions of rotations that are relevant to materials science. First, the probability distribution of orientations of volume elements in a single phase material is naturally expressed as a probability distribution of rotations from a reference orientation. While the properties of such functions are well-established \cite{1982bunge,2009mason_c}, this is a useful point of departure. The second part of the section follows from the observation that the macroscopic state of a grain boundary can be described by the orientations of the two grains adjoining a fixed interface \cite{2009olmsted_a}. A probability distribution of grain boundaries can then be written as a joint probability distribution of two three-dimensional rotations. Since the ability to independently transform the two constituent rotations is desired, this function is expressed as a probability distribution of four-dimensional rotations---a pair of three-dimensional rotations effectively describe a single four-dimensional rotation.

Section \ref{sec:grain_boundaries} begins by writing a probability distribution of grain boundaries as a probability distribution of four-dimensional rotations. Two different parameterizations of the basis functions used to expand such functions are then developed, one based on the description of a grain boundary by two boundary plane normals and a relative twist, and the other on the description by a misorientation and a boundary plane normal. Finally, the transformation properties of the basis functions with respect to three-dimensional rotations is described. With the basis functions defined, Section \ref{sec:symmetry} explores the various symmetries that a probability distribution of grain boundaries could obey, and develops a procedure to construct symmetrized basis functions. Several examples showing the efficacy of the symmetrization procedure are provided.

\section{Functions of rotations}
\label{sec:fun_of_rotations}

\subsection{Functions on SO(3)}
\label{subsec:func_so3}

Consider a set of objects $A$ with orientations described by some probability distribution on the space of rotations $g$ from a reference orientation. Provided that a complete orthogonal basis for functions on the space of rotations can be found, this probability distribution can be conveniently expressed as a linear combination of the basis functions. One consequence of the Peter--Weyl theorem is that the matrix elements of the irreducible unitary representations of a toplogical group provide such a basis for the expansion of functions on the group. Let the matrix elements of the $(2 a + 1)$-dimensional irreducible representation (irrep) of \su{2} be indicated by $U^{a}_{\alpha' \alpha}(g)$ in the standard basis \cite{1986altmann}, where $g$ has some suitable parameterization; these matrix elements form a complete orthogonal basis for functions on \su{2}. Moreover, the irreps of \so{3} are precisely those of \su{2} with $a \in \mathbb{N}$ (the nonnegative integers), meaning that this subset of matrix elements forms a complete orthogonal basis for functions on \so{3}. Finally, the hyperspherical harmonics $Z^{n}_{l m}(g)$ with $n \in 2 \mathbb{N}$ can be constructed from the matrix elements of the irreps of \so{3} by a multiple of a unitary transformation as in Eq.\ \ref{eq:z_to_su2}, and therefore form a complete orthogonal basis as well.

Suppose the orientations of the objects in $A$ are transformed by applying some rotation $g_r$ before $g$ and some rotation $g_l$ after $g$. This induces a transformation of the probability distribution of orientations; specifically, the transformed probability distribution can be written as the same linear combination of transformed hyperspherical harmonics. Since the composition of rotations is performed from right to left (discussed in Appendix \ref{app:convention}), the addition theorem in Eq.\ \ref{eq:su2_addition} allows the transformation of the matrix elements of the irreps of \so{3} to be written as
\begin{equation}
\sum_{\beta' \beta} U^a_{\alpha' \beta'}(g^{-1}_l) U^a_{\beta' \beta}(g) U^a_{\beta \alpha}(g^{-1}_r) = U^a_{\alpha' \alpha}(g^{-1}_l g g^{-1}_r)
\label{eq:so3_left_right}
\end{equation}
where the argument of the basis functions transforms in the dual sense to objects in $A$ to give the correct transformed probability distribution \footnote{Let $f(g)$ be a function of rotations, and define $f'(g) = f(g^{-1}_l g g^{-1}_r)$. Then the value of $f'(g_l g g_r)$ is equal to the value of $f(g)$.}. Using Eq.\ \ref{eq:su2_to_z} to write the matrix elements $U^a_{\beta' \beta}(g)$ and $U^a_{\alpha' \alpha}(g^{-1}_l g g^{-1}_r)$ as linear combinations of the hyperspherical harmonics gives
\begin{equation*}
\sum_{d \delta} \sum_{\beta' \beta} U^a_{\alpha' \beta'}(g^{-1}_l) \frac{\Pi_{d}}{\Pi^2_{a}} C^{a \beta}_{a \beta' d \delta} Z^{2 a}_{d \delta}(g) U^a_{\beta \alpha}(g^{-1}_r) = \sum_{c \gamma} \frac{\Pi_{c}}{\Pi^2_{a}} C^{a \alpha}_{a \alpha' c \gamma} Z^{2 a}_{c \gamma}(g^{-1}_l g g^{-1}_r)
\end{equation*}
where $C^{a \beta}_{a \beta' d \delta}$ is a Clebsch--Gordan coefficient and the symbol $\Pi_{a \dots b}$ stands for the product $[(2 a + 1) \dots (2 b + 1)]^{1/2}$. Multiplying through by $C^{a \alpha}_{a \alpha' e \epsilon}$, summing over $\alpha'$ and $\alpha$, and rearranging the remaining quantities yields
\begin{equation*}
\sum_{d \delta} Z^{2 a}_{d \delta}(g) \sum_{\mathclap{\alpha' \beta' \alpha \beta}} \frac{\Pi_{d e}}{\Pi^2_{a}} C^{a \beta}_{a \beta' d \delta} U^a_{\alpha' \beta'}(g^{-1}_l) U^a_{\beta \alpha}(g^{-1}_r) C^{a \alpha}_{a \alpha' e \epsilon} = Z^{2 a}_{e \epsilon}(g^{-1}_l g g^{-1}_r).
\end{equation*}
Since the irreps of \so{4} (the group of proper rotations in four dimensions) can be constructed from two irreps of \su{2} by \cite{1959racah,1961biedenharn,1967hughes}
\begin{equation}
R^{a b}_{c \gamma d \delta}(g_a, g_b) = \sum_{\mathclap{\alpha' \beta' \alpha \beta}} \frac{\Pi_{c d}}{\Pi^2_{b}} C^{b \beta'}_{a \alpha c \gamma} U^a_{\alpha' \alpha}(g^{-1}_a)  U^b_{\beta' \beta}(g^{-1}_b) C^{b \beta}_{a \alpha' d \delta},
\label{eq:so4_matrix}
\end{equation}
the action of \so{3} on a probability distribution on \so{3} can be written as an action of \so{4} on the hyperspherical harmonic basis for the expansion. That this transformation of the hyperspherical harmonics belongs to \so{4} should be expected; since the hyperspherical harmonics provide a complete orthogonal basis for functions on \su{2}, and \su{2} is isomorphic to the group of quaternions on $S^3$ (the unit sphere in four dimensions), the hyperspherical harmonics provide a complete orthogonal basis for functions on $S^3$ as well. More explicitly, the transformation rule for the hyperspherical harmonics is
\begin{equation*}
\sum_{d \delta} Z^{2 a}_{d \delta}(g) R^{a a}_{d \delta e \epsilon}(g_l, g_r) = Z^{2 a}_{e \epsilon}(g^{-1}_l g g^{-1}_r).
\end{equation*}
Notice that while Eq.\ \ref{eq:so4_matrix} allows the construction of irreps of \so{4} for all $a, b \in \mathbb{N}$, the requirement that the irreps in Eq.\ \ref{eq:so3_left_right} have the same dimension places the restriction $a = b$ on those that transform the hyperspherical harmonics \cite{1961biedenharn,1967hughes}.

\subsection{Functions on SO(4)}
\label{subsec:func_so4}

Consider partitioning the set of objects into two subsets $A$ and $B$, and requiring that the orientations of the objects in $A$ and $B$ transform independently. Since the orientations of the objects in a single set can be described by a probability distribution on \so{3}, the orientations of the objects in $A$ and $B$ can be described by a joint probability distribution on a space with the necessary transformation properties. Invoking the Peter--Weyl theorem as in Section \ref{subsec:func_so3}, the quantities defined by
\begin{align}
\hat{R}^{a b}_{\alpha' \beta' \alpha \beta}(g_a, g_b) &= \sum_{\mathclap{c \gamma d \delta}} \frac{\Pi_{a c d}}{2 \pi^2 \Pi_{b}} C^{b \beta'}_{a \alpha c \gamma} R^{a b}_{c \gamma d \delta}(g_a, g_b) C^{b \beta}_{a \alpha' d \delta} \nonumber \\
&= \frac{\Pi_{a b}}{2 \pi^2} U^{a}_{\alpha' \alpha}(g^{-1}_a) U^{b}_{\beta' \beta}(g^{-1}_b)
\label{eq:so4_hat}
\end{align}
are related to the irreps of \so{4} in Eq.\ \ref{eq:so4_matrix} by a multiple of a unitary transformation, and therefore form a complete orthogonal basis for functions on \so{4}. The orthogonality relation for the $R^{a b}_{c \gamma d \delta}(g_a, g_b)$ in Eq.\ \ref{eq:so4_orthogonal} further implies that this basis is orthonormal. Since $\so{3} \times \so{3}$ is a quotient group of \so{4}, this basis can be used to expand a joint probability distribution of orientations. The transformation properties of this basis can be derived from the addition theorem for the irreps of \so{3} as follows:
\begin{align*}
\sum_{\mathclap{\gamma' \delta' \gamma \delta}} \frac{4 \pi^4}{\Pi^2_{a b}} \hat{R}^{a b}_{\alpha' \beta' \gamma' \delta'}(g^{-1}_1, g^{-1}_2) \hat{R}^{a b}_{\gamma' \delta' \gamma \delta}(g_3, g_4) \hat{R}^{a b}_{\gamma \delta \alpha \beta}(g^{-1}_5, g^{-1}_6) = \sum_{\mathclap{\gamma' \delta' \gamma \delta}} \frac{\Pi_{a b}}{2 \pi^2} & U^{a}_{\alpha' \gamma'}(g_1) U^{a}_{\gamma' \gamma}(g^{-1}_3) U^{a}_{\gamma \alpha}(g_5) \\ & U^{b}_{\beta' \delta'}(g_2) U^{b}_{\delta' \delta}(g^{-1}_4) U^{b}_{\delta \beta}(g_6).
\end{align*}
The irreps of \so{3} can be multiplied with Eq.\ \ref{eq:su2_addition} and the result written as one of the basis functions in Eq.\ \ref{eq:so4_hat} to find
\begin{equation*}
\sum_{\mathclap{\gamma' \delta' \gamma \delta}} \frac{4 \pi^4}{\Pi^2_{a b}} \hat{R}^{a b}_{\alpha' \beta' \gamma' \delta'}(g^{-1}_1, g^{-1}_2) \hat{R}^{a b}_{\gamma' \delta' \gamma \delta}(g_3, g_4) \hat{R}^{a b}_{\gamma \delta \alpha \beta}(g^{-1}_5, g^{-1}_6) = \hat{R}^{ab}_{\alpha' \beta' \alpha \beta}(g^{-1}_5 g_3 g^{-1}_1, g^{-1}_6 g_4 g^{-1}_2).
\end{equation*}
Interpreting $g_3$ and $g_4$ as variables and $g_1$, $g_2$, $g_5$ and $g_6$ as parameters, this shows that the rotations $g_3$ and $g_4$ can be transformed independently. Moreover, expanding the left side allows this transformation to be written as right multiplication of the row vector of basis functions by a transformation matrix:
\begin{equation}
\sum_{\mathclap{\gamma' \delta' \gamma \delta}} \hat{R}^{a b}_{\gamma' \delta' \gamma \delta}(g_3, g_4) U^{a}_{\alpha' \gamma'}(g_1) U^{b}_{\beta' \delta'}(g_2) U^{a}_{\gamma \alpha}(g_5) U^{b}_{\delta \beta}(g_6) = \hat{R}^{ab}_{\alpha' \beta' \alpha \beta}(g^{-1}_5 g_3 g^{-1}_1, g^{-1}_6 g_4 g^{-1}_2).
\label{eq:so4_transform}
\end{equation}
That is, if the joint probability distribution of the orientations of objects in $A$ and $B$ is expanded using the basis functions in Eq.\ \ref{eq:so4_hat}, Eq.\ \ref{eq:so4_transform} transforms the basis functions in a way that is consistent with independent transformations of the orientations of objects in $A$ and $B$ (similar to Eq.\ \ref{eq:so3_left_right}, the arguments transform in the dual sense to the objects).

\section{Grain boundary parameterizations}
\label{sec:grain_boundaries}

The available literature on the topology and metric of the space of homophase grain boundaries is particularly concerned with the space around the null boundary. More specifically, the five available dimensions are often separated into three that identify the misorientation of the grains and two that identify the boundary plane normal. When the misorientation of the grains corresponds to the identity element there is no identifiable grain boundary, and the two dimensions that identify the boundary plane appear to be redundant. Many of the existing approaches contract this null boundary subspace to a single point, and perform some ad hoc deformation of the surrounding space and metric \cite{2006cahn,2009morawiec}.

The view of the authors is instead that the null boundary subspace should remain, and the values of certain functions be constrained to be uniform on the null boundary subspace. This position is justified by two observations. First, the points of the null boundary subspace all correspond to distinct heterophase grain boundaries; contracting the null boundary subspace necessarily restricts the theory to only homophase grain boundaries. Second, it is not clear that all functions should be single-valued on the null boundary subspace; consider that the mobility of a low-angle homophase boundary is expected to be a strong function of the Burgers vectors of the constituent dislocations. Several trajectories can be constructed in the homophase grain boundary space along which the misorientation angle and dislocation density continuously decrease, but the Burgers vectors of the dislocations remain constant. The limiting values of the mobility along these trajectories as they approach the null boundary subspace are not necessarily the same, contradicting one of the initial motivations for contracting the null boundary subspace.

Related considerations have already been applied for orientation distribution functions. The crystal point group symmetries can be enforced either by restricting the space of orientations to a single fundamental zone, or by restricting the functions on the space to remain invariant to the action of the point group symmetry. The latter has the advantage that the underlying space is always \so{3} and does not change with the point group symmetry of the crystals, avoiding the difficulty of parameterizing a space with complicated topology.

Our approach to construct a metric and suitable basis functions on the space of grain boundaries begins with the case of least symmetry (heterophase boundaries and triclinic crystals), and gradually introduces symmetries by placing constraints on the allowed linear combinations of the basis functions. Let a grain boundary be approximated by some collection of planar patches. Given a patch, a coordinate system is constructed that embeds the patch in the $x$-$y$ plane. The grain containing the negative $z$ axis is identified as grain one, and the grain containing the positive $z$ axis as grain two. Let $g_1$ and $g_2$ be rotations that bring a grain initially aligned with the coordinate system into coincidence with grains one and two, respectively. These rotations completely describe the macroscopic state of the grain boundary patch, though with one more variable than is necessary. This follows from the ambiguity in the construction of the coordinate system; observe that applying the same rotation about the $z$ axis to grains one and two changes $g_1$ and $g_2$, but not the macroscopic state of the grain boundary patch. This suggests that $g_1$ and $g_2$ be written as products of simpler rotations that separate the dependence on the shared rotation about the $z$ axis.

\subsection{NNT parameterization}
\label{subsec:NNT_param}

\begin{figure}
    \centering
    \includegraphics[width=9cm]{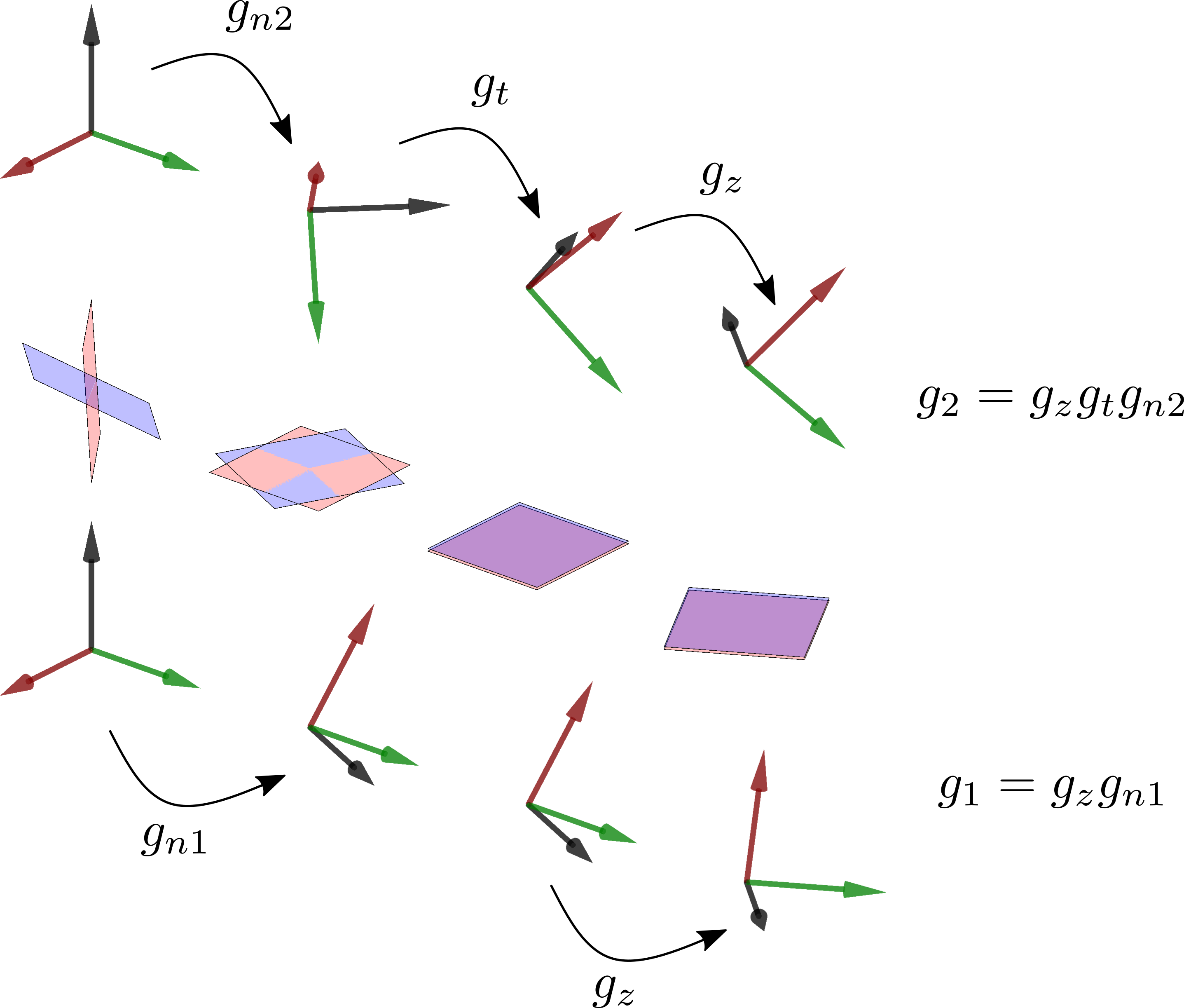}
    \caption{Schematic illustrating the construction of a grain boundary using the conventions established in the normal-normal-twist (NNT) parameterization.}
    \label{fig:nnt}
\end{figure}

Let $g_{n1}$ and $g_{n2}$ be rotations about axes in the $x$-$y$ plane, and $g_t$ and $g_z$ be rotations about the $z$ axis. The normal-normal-twist (NNT) parameterization gives the orientations of grains one and two as $g_1 = g_z g_{n1}$ and $g_2 = g_z g_t g_{n2}$, respectively. Observe that $g_{n1}$ and $g_{n2}$ set the crystallographic boundary planes of the two grains, that $g_t$ is a rotation of grain two relative to grain one about the $z$ axis, and that $g_z$ is a shared rotation about the $z$ axis. The steps involved in constructing such a bicrystal with the grain boundary in the $x$-$y$ plane are illustrated in Figure \ref{fig:nnt}.

Let $\omega_{n1}$ and $\phi_{n1}$ be the rotation and azimuthal angles of $g_{n1}$, $\omega_{n2}$ and $\phi_{n2}$ be the rotation and azimuthal angles of $g_{n2}$, and $\omega_t$ and $\omega_z$ be the rotation angles of $g_t$ and $g_z$, respectively. A metric on the grain boundary space can be found by observing that \su{2} is isomorphic to the group of unit quaternions on $S^3$. Hence, a metric on $\su{2} \times \su{2}$ is induced by the embedding of $S^3 \times S^3$ in $\mathbb{R}^8$ with the Euclidean metric. The result of the calculation in App.\ \ref{app:NNT_metric} is that the differential volume element on $\su{2} \times \su{2}$ is
\begin{equation*}
\diff \Omega = \frac{1}{64} \abs{ \sin \omega_{n1} } \abs{ \sin \omega_{n2} } \diff \omega_{n1} \diff \phi_{n1} \diff \omega_{n2} \diff \phi_{n2} \diff \omega_t \diff \omega_z
\end{equation*}
in the NNT parameterization. Integrating over $\omega_z$ will subsequently give the differential volume element on the grain boundary space.

The requirement that the orientations of grains one and two transform independently (e.g., by point group symmetries) suggests that the basis functions be constructed from the ones for \so{4}. Specifically, expanding one of the basis functions in Eq.\ \ref{eq:so4_hat} and separating the constituent rotations with Eq.\ \ref{eq:su2_addition} gives
\begin{equation*}
\hat{R}^{a b}_{\alpha' \beta' \alpha \beta}(g_1, g_2) = \sum_{\mathclap{\gamma \delta' \delta}} \frac{\Pi_{a b}}{2 \pi^2} U^{a}_{\alpha' \gamma}(g^{-1}_{n1}) U^{a}_{\gamma \alpha}(g^{-1}_{z}) U^{b}_{\beta' \delta'}(g^{-1}_{n2}) U^{b}_{\delta' \delta}(g^{-1}_t) U^{b}_{\delta \beta}(g^{-1}_{z}).
\end{equation*}
The matrix elements of the irreps of \so{3} for a rotation about the $z$ axis have the particularly convenient form $U^{a}_{\alpha' \alpha}(g_z) = \delta_{\alpha' \alpha} e^{-i \alpha \omega}$, where $\omega$ is the angle of rotation. Making this substitution for all matrix elements that depend on $g_z$ and $g_t$ yields
\begin{equation}
\hat{R}^{a b}_{\alpha' \beta' \alpha \beta}(g_1, g_2) = \frac{\Pi_{a b}}{2 \pi^2} U^{a}_{\alpha' \alpha}(g^{-1}_{n1}) U^{b}_{\beta' \beta}(g^{-1}_{n2}) e^{i \beta \omega_{t}} e^{i (\alpha + \beta) \omega_{z}}.
\label{eq:NNT_middle}
\end{equation}
This leaves the matrix elements of the irreps of \so{3} for the rotations $g_{n1}$ and $g_{n2}$. A rotation by the angle $\omega$ about an axis with the spherical angles $(\pi / 2, \phi)$ is equivalent to a rotation about the $z$ axis by $-\phi + \pi / 2$, a rotation about the $y$ axis by $\omega$, and a rotation about the $z$ axis by $\phi - \pi / 2$:
\begin{equation*}
U^{a}_{\alpha' \alpha}(\omega, \pi / 2, \phi) = \sum_{\beta' \beta} U^{a}_{\alpha' \beta'}(\phi - \pi / 2, 0, 0) U^{a}_{\beta' \beta}(\omega, \pi / 2, \pi / 2) U^{a}_{\beta \alpha}(-\phi + \pi / 2, 0, 0)
\end{equation*}
The matrix element $U^{a}_{\beta' \beta}(\omega, \pi / 2, \pi / 2)$ is known as the Wigner little $d$-function $d^{a}_{\beta' \beta}(\omega)$ \cite{1988varshalovich}. Making the same substitution as above for the rotations about the $z$ axis gives
\begin{equation}
U^{a}_{\alpha' \alpha}(\omega, \pi / 2, \phi) = i^{\alpha' - \alpha} d^{a}_{\alpha' \alpha}(\omega) e^{-i (\alpha' - \alpha) \phi}
\label{eq:little_d}
\end{equation}
for an arbitrary rotation about an axis in the $x$-$y$ plane. Using this to substitute for the remaining matrix elements of the irreps of \so{3} in Eq.\ \ref{eq:NNT_middle} and using the symmetry relation $d^{a}_{\alpha' \alpha}(-\omega) = (-1)^{\alpha' - \alpha} d^{a}_{\alpha' \alpha}(\omega)$ of the little $d$-function yields
\begin{equation}
\hat{R}^{a b}_{\alpha' \beta' \alpha \beta}(g_1, g_2) = \frac{\Pi_{a b}}{2 \pi^2} (-i)^{\alpha' - \alpha + \beta' - \beta} d^{a}_{\alpha' \alpha}(\omega_{n1}) d^{b}_{\beta' \beta}(\omega_{n2}) e^{-i (\alpha' - \alpha) \phi_{n1}} e^{-i (\beta' - \beta) \phi_{n2}} e^{i \beta \omega_{t}} e^{i (\alpha + \beta) \omega_{z}}.
\label{eq:NNT_full}
\end{equation}
These basis functions are entirely equivalent to those in Eq.\ \ref{eq:so4_hat}, the only difference being a choice of parameterization; the practitioner can substitute one for the other depending on the need for computational convenience or interpretability.

As discussed in the introduction to this section, one symmetry that all functions on the grain boundary space should obey is invariance to the shared rotation about the $z$ axis, i.e., the value of $\omega_z$. By inspection, the basis functions with this property satisfy the condition $\alpha = -\beta$. The basis functions on the grain boundary space in the NNT parameterization can then be written as
\begin{equation}
N^{a b}_{\alpha \beta \gamma}(\omega_{n1}, \phi_{n1}, \omega_{n2}, \phi_{n2}, \omega_{t}) = \frac{\Pi_{a b}}{\sqrt{2 \pi^3}} (-i)^{\alpha + \beta} d^{a}_{\alpha \gamma}(\omega_{n1}) d^{b}_{\beta -\gamma}(\omega_{n2}) e^{-i (\alpha - \gamma) \phi_{n1}} e^{-i (\beta + \gamma) \phi_{n2}} e^{-i \gamma \omega_{t}}
\label{eq:NNT_reduced}
\end{equation}
where $a, b \in \mathbb{N}$, $\alpha \in [-a, a]$, $\beta \in [-b, b]$, and $\gamma \in [-a, a] \cap [-b, b]$. The corresponding volume element on the grain boundary space is
\begin{equation*}
\diff \Omega = \frac{1}{64} \abs{ \sin \omega_{n1} } \abs{ \sin \omega_{n2} } \diff \omega_{n1} \diff \phi_{n1} \diff \omega_{n2} \diff \phi_{n2} \diff \omega_t
\end{equation*}
where the allowed intervals of the five variables are $[0, 2 \pi]$, $[0, 2 \pi]$, $[0, 2 \pi]$, $[0, 2 \pi]$, and $[0, 2 \pi]$. The action of point group symmetries on this basis will be discussed in Section \ref{subsec:rotations_of_gbs}.

\subsection{MBP parameterization}
\label{subsec:MBP_param}

\begin{figure}
    \centering
    \includegraphics[width=9cm]{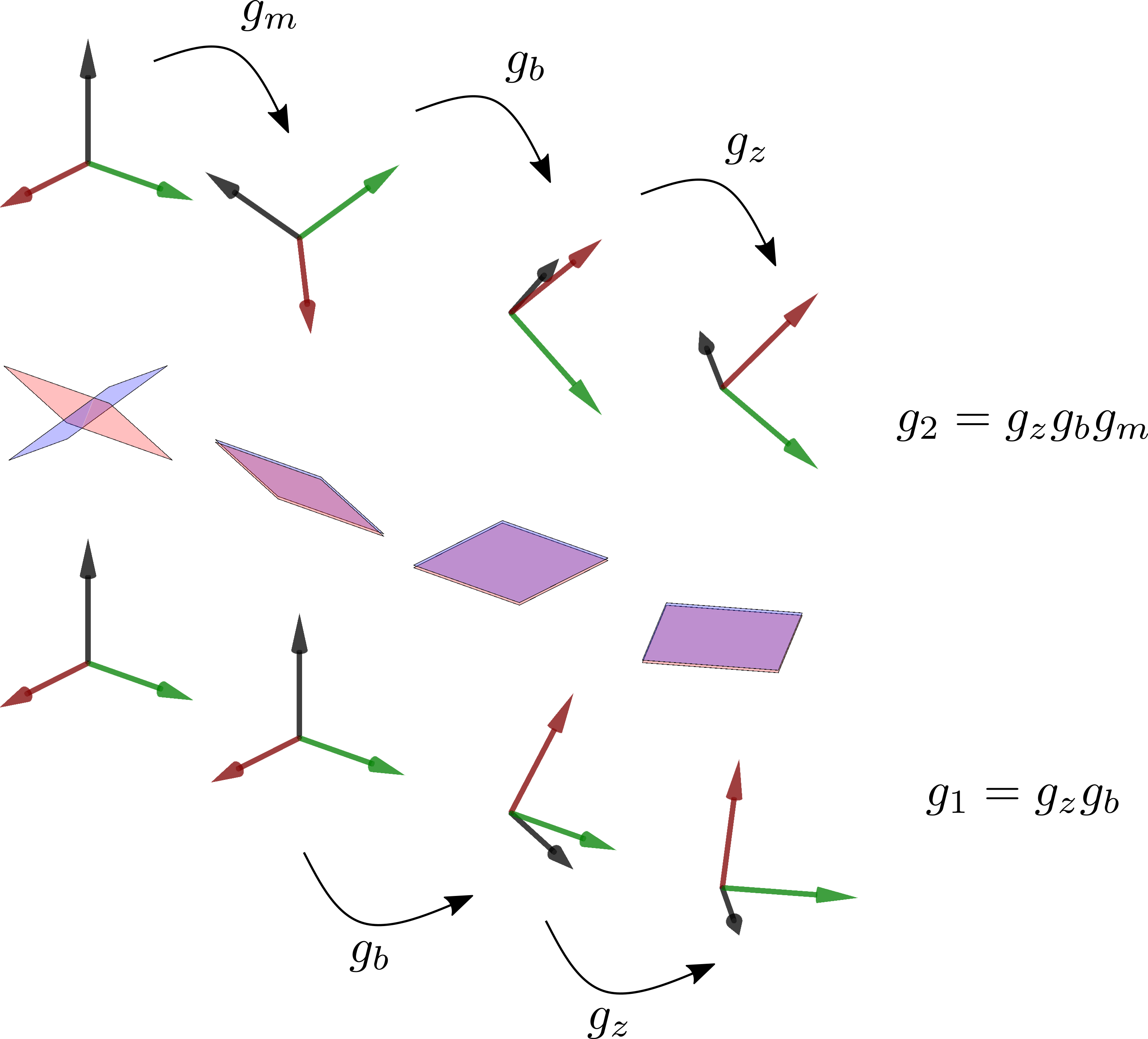}
    \caption{Schematic illustrating the construction of a grain boundary using the conventions established in the misorientation-boundary-plane (MBP) parameterization.}
    \label{fig:mbp}
\end{figure}

Let $g_{m}$ be an arbitrary rotation, $g_{b}$ be a rotation about an axis in the $x$-$y$ plane, and $g_z$ be a rotation about the $z$ axis. The misorientation-boundary-plane (MBP) parameterization gives the orientations of grains one and two as $g_1 = g_z g_b$ and $g_2 = g_z g_b g_m$, respectively. Observe that $g_m$ is the misorientation of grain two relative to grain one, that $g_b$ is a shared rotation that sets the boundary plane of grain one, and that $g_z$ is a shared rotation about the $z$ axis. The steps involved in constructing such a bicrystal with the grain boundary in the $x$-$y$ plane are illustrated in Figure \ref{fig:mbp}.

Let $\omega_m$, $\theta_m$, and $\phi_m$ be the rotation, polar and azimuthal angles of $g_m$, $\omega_b$ and $\phi_b$ be the rotation and azimuthal angles of $g_b$, and $\omega_z$ be the rotation angle of $g_z$. As for the NNT parameterization in Sec.\ \ref{subsec:NNT_param}, a metric on the grain boundary space is induced by embedding $S^3 \times S^3$ in $\mathbb{R}^8$ with the Euclidean metric. The result of the calculation in App.\ \ref{app:MBP_metric} is that the differential volume element on $\su{2} \times \su{2}$ is
\begin{equation*}
\diff \Omega = \frac{1}{16} \sin^2(\omega_m / 2) \abs{\sin \theta_m} \abs{\sin \omega_b} \diff \omega_m \diff \theta_m \diff \phi_m \diff \omega_b \diff \phi_b \diff \omega_z
\end{equation*}
in the MBP parameterization. Integrating over the variable $\omega_z$ will subsequently give the differential volume element on the grain boundary space.

As before, the requirement that the orientations of the grains transform independently suggests that the basis functions be constructed from the ones for \so{4}. Expanding one of the basis functions given by Eq.\ \ref{eq:so4_hat} and separating the constituent rotations with Eq.\ \ref{eq:su2_addition} gives
\begin{equation*}
\hat{R}^{a b}_{\alpha' \beta' \alpha \beta}(g_1, g_2) = \sum_{\mathclap{\gamma \delta' \delta}} \frac{\Pi_{a b}}{2 \pi^2} U^{a}_{\alpha' \gamma}(g^{-1}_b) U^{a}_{\gamma \alpha}(g^{-1}_z) U^{b}_{\beta' \delta'}(g^{-1}_m) U^{b}_{\delta' \delta}(g^{-1}_b) U^{b}_{\delta \beta}(g^{-1}_z).
\end{equation*}
Substituting for all matrix elements that are functions of $g_z$ and using the Clebsch--Gordan expansion in Eq.\ \ref{eq:clebsh_gordan} to expand the direct product of functions of $g_b$ yields
\begin{equation*}
\hat{R}^{a b}_{\alpha' \beta' \alpha \beta}(g_1, g_2) = \sum_{e} \sum_{\delta' \epsilon' \epsilon} \frac{\Pi_{a b}}{2 \pi^2} C^{e \epsilon'}_{a \alpha' b \delta'} C^{e \epsilon}_{a \alpha b \beta} U^{b}_{\beta' \delta'}(g^{-1}_m) U^{e}_{\epsilon' \epsilon}(g^{-1}_b) e^{i (\alpha + \beta) \omega_z}.
\end{equation*}
The matrix element depending on $g_m$ is expanded with Eq.\ \ref{eq:su2_to_z}, the argument inverted with Eq.\ \ref{eq:invert_z}, and the matrix element depending on $g_b$ written as a little $d$-function using Eq. \ref{eq:little_d} to find
\begin{align}
\hat{R}^{a b}_{\alpha' \beta' \alpha \beta}(g_1, g_2) = \sum_{e f} \sum_{\delta' \epsilon' \epsilon \phi} & \frac{\Pi_{a f}}{\sqrt{2} \pi \Pi_{b}} (-i)^{\epsilon' - \epsilon} C^{e \epsilon'}_{a \alpha' b \delta'} C^{b \delta'}_{f \phi b \beta'} C^{e \epsilon}_{a \alpha b \beta} \nonumber \\
& Z^{2 b}_{f \phi}(g_m) d^{e}_{\epsilon' \epsilon}(\omega_b) e^{-i (\epsilon' - \epsilon) \phi_b} e^{i (\alpha + \beta) \omega_z}.
\label{eq:MBP_full}
\end{align}
As before, these basis functions are entirely equivalent to those in Eqs.\ \ref{eq:so4_hat} and \ref{eq:NNT_full} except for the parameterization. Of the three, this form is likely the most expensive to evaluate, but offers the closest correspondence to the conventional description of a grain boundary by a misorientation and a boundary plane.

Given the equivalence of Eqs.\ \ref{eq:NNT_full} and \ref{eq:MBP_full}, invarance to the shared rotation about the $z$ axis is again enforced by the condition $\alpha = -\beta$. The properties of the Clebsch--Gordan coefficients then require that $\epsilon$ equal zero, or after relabeling the indices:
\begin{equation*}
M^{a b}_{\alpha \beta \gamma}(\omega_m, \theta_m, \phi_m, \omega_b, \phi_b) = \sum_{e f} \sum_{\delta \epsilon \phi} \frac{\Pi_{a f}}{\sqrt{\pi} \Pi_{b}} (-i)^{\epsilon} C^{e \epsilon}_{a \alpha b \delta} C^{b \delta}_{f \phi b \beta} C^{e 0}_{a \gamma b -\gamma} Z^{2 b}_{f \phi}(g_m) d^{e}_{\epsilon 0}(\omega_b) e^{-i \epsilon \phi_b}.
\end{equation*}
Finally, the sign of $\epsilon$ is inverted and the little $d$-function and complex exponential written as a spherical harmonic using Eq.\ \ref{eq:spherical_harmonic} to find
\begin{equation}
M^{a b}_{\alpha \beta \gamma}(\omega_m, \theta_m, \phi_m, \omega_b, \phi_b) = \sum_{e f} \sum_{\delta \epsilon \phi} \frac{2 \Pi_{a f}}{\Pi_{b e}} (-i)^{\epsilon} C^{e -\epsilon}_{a \alpha b \delta} C^{b \delta}_{f \phi b \beta} C^{e 0}_{a \gamma b -\gamma} Z^{2 b}_{f \phi}(g_m) Y^{\epsilon}_{e}(\omega_b, \phi_b)
\label{eq:MBP_reduced}
\end{equation}
as the basis functions on the grain boundary space in the MBP parameterization. The corresponding volume element on the grain boundary space is
\begin{equation*}
\diff \Omega = \frac{1}{16} \sin^2(\omega_m / 2) \abs{\sin \theta_m} \abs{\sin \omega_b} \diff \omega_m \diff \theta_m \diff \phi_m \diff \omega_b \diff \phi_b
\end{equation*}
where the allowed intervals of the five variables are $[0, 2 \pi]$, $[0, \pi]$, $[0, 2 \pi]$, $[0, 2 \pi]$ and $[0, 2 \pi]$. When performing integrals over this space, it is useful to observe that the domain of the spherical harmonics is twice the usual domain, and that Eq.\ \ref{eq:clebsh_gordan_sum} is a special case of one of the standard sums involving a product of two Clebsh--Gordan coefficients.

\subsection{Rotations of the grain boundary space}
\label{subsec:rotations_of_gbs}

Practically speaking, the main rotations of concern for functions on the grain boundary space are those belonging to the crystallographic point groups. Given the conventions established in Appendix \ref{app:convention}, these symmetry operations should be performed before the rotations that establish the grain boundary. That is, if $g_{s1}$ and $g_{s2}$ are symmetry operations belonging to the crystallographic point groups of the respective grains, then the required transformation of the $\hat{R}^{a b}_{\gamma' \delta' \gamma \delta}(g_1, g_2)$ is given by Eq.\ \ref{eq:so4_transform} as
\begin{equation*}
\sum_{\mathclap{\gamma' \delta' \gamma \delta}} \hat{R}^{a b}_{\gamma' \delta' \gamma \delta}(g_1, g_2) U^{a}_{\alpha' \gamma'}(g_{s1}) U^{b}_{\beta' \delta'}(g_{s2}) U^{a}_{\gamma \alpha}(E) U^{b}_{\delta \beta}(E) = \hat{R}^{a b}_{\alpha' \beta' \alpha \beta}(g_1 g^{-1}_{s1}, g_2 g^{-1}_{s2})
\end{equation*}
where $E$ is the identity rotation. Making the substitution $U^{a}_{\alpha' \alpha}(E) = \delta_{\alpha' \alpha}$ for the identity rotations and performing the summations over $\gamma$ and $\delta$ gives
\begin{equation*}
\sum_{\mathclap{\gamma' \delta'}} \hat{R}^{a b}_{\gamma' \delta' \alpha \beta}(g_1, g_2) U^{a}_{\alpha' \gamma'}(g_{s1}) U^{b}_{\beta' \delta'}(g_{s2}) = \hat{R}^{a b}_{\alpha' \beta' \alpha \beta}(g_1 g^{-1}_{s1}, g_2 g^{-1}_{s2}).
\end{equation*}
Restricting this to the basis functions that satisfy the condition $\alpha = -\beta$ and relabeling the indices then yields the required transformation of the basis functions on the grain boundary space. This is written using the notation for, e.g., the MBP parameterization, as
\begin{equation}
\sum_{\mathclap{\delta \epsilon}} M^{a b}_{\delta \epsilon \gamma}(g_1, g_2) U^{a}_{\alpha \delta}(g_{s1}) U^{b}_{\beta \epsilon}(g_{s2}) = M^{a b}_{\alpha \beta \gamma}(g_1 g^{-1}_{s1}, g_2 g^{-1}_{s2}).
\label{eq:gb_rotations}
\end{equation}
If this is interpreted as right multiplication of a row vector of basis functions by a transformation matrix, then the transformation matrix can be formed from the Kroneker product of two transposed irreps of \so{3} and one identity matrix.

Perhaps defying expectations, the application of a symmetry operation involving rotations performed after those that establish the grain boundary is considerably more difficult. Beginning with Eq.\ \ref{eq:so4_transform} and performing analogous steps to those above gives
\begin{equation*}
\sum_{\mathclap{\delta \epsilon}} \hat{R}^{a b}_{\alpha \beta \delta \epsilon}(g_1, g_2) U^{a}_{\delta \gamma}(g_{s1}) U^{b}_{\epsilon -\gamma}(g_{s2}) = M^{ab}_{\alpha \beta \gamma}(g^{-1}_{s1} g_1, g^{-1}_{s2} g_2)
\end{equation*}
for general symmetry operations $g_{s1}$ and $g_{s2}$. That is, the construction of the transformed basis functions generally involves all of the elements of the irreps of \so{4} and cannot be written as right multiplication of a row vector of basis functions by a transformation matrix. Simplification is possible in certain cases though. Specifically, if $g_{s1}$ and $g_{s2}$ are rotations by $\pi$ about the $y$ axis, then $U^{a}_{\alpha' \alpha}(g^{-1}_{\pi y}) = (-1)^{a - \alpha} \delta_{\alpha' -\alpha}$ and the above equation reduces to
\begin{equation}
(-1)^{a + b} M^{a b}_{\alpha \beta -\gamma}(g_1, g_2) = M^{ab}_{\alpha \beta \gamma}(g_{\pi y} g_1, g_{\pi y} g_2).
\label{eq:gb_rotations_l}
\end{equation}
This can be written as right multiplication of a row vector of basis functions by a transformation matrix, and will be useful in Section \ref{subsec:point_group}.

\subsection{Numerical considerations}
\label{subsec:numerical}

By design, the basis functions on the grain boundary space in the NNT parameterization and the basis functions on the grain boundary space in the MBP parameterization are precisely equivalent. A reasonable question would be what purpose is served by developing two different forms for the same set of functions. First, the flexibility to select among several parameterizations can help to relate a point in the grain boundary space to physically meaningful quantities. After all, the purpose of this exercise is to facilitate the communication and interpretation of the properties of some physical system. Second, integrating over some of the variables allows a specific subset of information to be retained. This can be useful for certain mathematical applications, including enforcing the null boundary symmetry in Section \ref{subsec:null_boundary}.

For numerical calculations though, the situation is quite different. Particularly for the MBP parameterization, calculating a basis function's value using Eq.\ \ref{eq:MBP_reduced} would involve evaluating and summing products of the hyperspherical and spherical harmonics and the various special functions that they contain. Since such a procedure could cause the accumulation of significant error, the authors instead suggest that the basis functions be evaluated using Eq.\ \ref{eq:so4_hat}. Specifically, given the rotations $g_1$ and $g_2$ in any parameterization, a basis function's value can be found with
\begin{equation}
M^{a b}_{\alpha \beta \gamma}(g_1, g_2) = \frac{\Pi_{a b}}{\sqrt{2 \pi^3}} U^{a}_{\alpha \gamma}(g^{-1}_1) U^{b}_{\beta -\gamma}(g^{-1}_2)
\label{eq:numerical}
\end{equation}
where the change in normalization is a consequence of the invariance of a grain boundary to shared rotations of the grains about the $z$ axis. If the necessary rotations are written using the Cayley--Klein parameters, then the irreps of \so{3} in this equation can be conveniently evaluated using, e.g., a formula on page 242 of Ref.\ \cite{1986altmann} which follows the same conventions as this article.

\section{Symmetries}
\label{sec:symmetry}

Let $f(g_1, g_2)$ be a function on \so{4} written as an expansion over the $\hat{R}^{a b}_{\alpha' \beta' \alpha \beta}(g_1, g_2)$ in any parameterization. Provided that $f(g_1, g_2)$ satisfies a few technical constraints (e.g., is of bounded variation), then this function can be written as the inner product
\begin{equation*}
f(g_1, g_2) = \langle \hat{R} | c \rangle
\end{equation*}
where $\langle \hat{R} |$ is a row vector of the basis functions, $| c \rangle$ is a column vector of the expansion coefficients, and equality indicates that the expansion converges everywhere. This expansion is not particularly efficient though. Any symmetries satisfied by the physical system restrict the allowed values of $| c \rangle$ to a smaller vector subspace, and a function that satisfies these symmetries is more concisely expanded over the linear combinations of the $\hat{R}^{a b}_{\alpha' \beta' \alpha \beta}(g_1, g_2)$ that form a basis for the smaller subspace. This is particularly useful when constructing a probability distribution from some sampled population (e.g., a grain boundary character distribution) since enforcing symmetries can reduce the number of samples necessary for a given uncertainty by several orders of magnitude.

For a physical system with $s$ symmetries, let $X_{i}$ be the vector subspace of coefficients $| c \rangle$ that satisfies the first to the $i$th symmetries, and $Y_{i + 1}$ be the vector subspace of coefficients that satisfies the $(i + 1)$th symmetry. The vector subspace $X_{i + 1}$ could then be constructed as the intersection of $X_{i}$ and $Y_{i + 1}$, and repeating the procedure would allow symmetries to be sequentially enforced until the set of symmetries is exhausted and $X_{s}$ is found. Given some matrix $\mat{X}_{s}$ with columns that form an orthonormal basis for $X_{s}$, a function $\mathring{f}(g_1, g_2)$ satisfying all of the symmetries could then be written as
\begin{align*}
\mathring{f}(g_1, g_2) &= \langle \hat{R} | \mat{X}_s \mat{X}^{\dagger}_s | c \rangle \\
&= \langle \mathring{R} | \mathring{c} \rangle
\end{align*}
where a dagger indicates the matrix adjoint, $\langle \mathring{R} | = \langle \hat{R} | \mat{X}_{s}$ is the set of symmetrized basis functions, and $| \mathring{c} \rangle = \mat{X}^{\dagger}_{s} | c \rangle$ is the set of symmetrized expansion coefficients. The nontrivial step in this line of reasoning is the construction of $X_{i + 1}$ given $X_{i}$ and $Y_{i + 1}$.

If $\mat{X}_{i}$ is a matrix with columns that form an orthonormal basis for $X_{i}$, and $\mat{A}_{i + 1}$ is any matrix that leaves invariant only the subspace $Y_{i + 1}$, then $\mat{X}_{i}^{\dagger} \mat{A}_{i + 1} \mat{X}_{i}$ is a matrix written in the basis $\mat{X}_{i}$ that leaves invariant only the intersection of $X_{i}$ and $Y_{i + 1}$. Furthermore, if $\mathrm{N}(\mat{B})$ is a matrix with columns that form an orthonormal basis for the nullspace of $\mat{B}$, then $\mathrm{N}(\mat{X}_{i}^{\dagger} \mat{A}_{i + 1} \mat{X}_{i} - \mat{I})$ is a basis for the intersection of $X_{i}$ and $Y_{i + 1}$ written in the basis $\mat{X}_{i}$. The purpose of using the basis $\mat{X}_{i}$ is that the dimension of $X_{i}$ is often considerably smaller than that of the ambient space, reducing the computational burden and the accumulated numerical error. The result is that an orthonormal basis $\mat{X}_{i + 1}$ for the subspace $X_{i + 1}$ can be found by
\begin{equation}
\mat{X}_{i + 1} = \mat{X}_{i} \mathrm{N}(\mat{X}_{i}^{\dagger} \mat{A}_{i + 1} \mat{X}_{i} - \mat{I})
\label{eq:add_symmetry}
\end{equation}
provided that a suitable matrix $\mat{A}_{i + 1}$ is known. The following sections describe the construction of these matrices for a variety of symmetry operations.

Practically speaking, the procedure to find the symmetrized basis functions on the grain boundary space is the same as that for the symmetrized basis functions on \so{4}, the only difference being that the rows of the $\mat{X}_{i}$ and $\mat{A}_{i + 1}$ are fewer in number. When constructing the required matrices, the practitioner is suggested to choose some maximum value of $a + b$, and then construct the $\mat{X}_{i}$ in blocks with increasing values of this quantity. A reasonable convention for the ordering of the rows within a block is given by arranging the labels $(a, b, \gamma, \alpha, \beta)$ of the basis functions on the grain boundary space in lexicographic order, where the relative positions of the lower indices facilitates the construction of the $\mat{A}_{i + 1}$. The resulting $\mat{X}_s$ would define the symmetrized basis functions for both the NNT and MBP parameterizations, since the functions defined in Eqs.\ \ref{eq:NNT_reduced} and \ref{eq:MBP_reduced} are identical by construction.

\subsection{Point group symmetry}
\label{subsec:point_group}

For a heterophase boundary, the crystallographic points groups of the grains adjoining a grain boundary can be different. Let $S_1$ and $S_2$ be sets of generators for the rotational point groups of grains one and two; then for any element in $(S_1 \times E) \cup (E \times S_2)$, Eq.\ \ref{eq:gb_rotations} defines a block diagonal matrix $\mat{A}$ that transforms the basis functions on the grain boundary space according to that generator. Reference to Eq.\ \ref{eq:gb_rotations} shows that the construction of $\mat{A}$ requires an explicit formula for the irreps of \so{3}; one convenient choice using the Cayley--Klein parameters is available on page 242 of Ref.\ \cite{1986altmann}. Finally, enforcing the symmetry of each of the elements of $(S_1 \times E) \cup (E \times S_2)$ individually is sufficient to enforce the entire rotational point groups of grains one and two.

If the crystallographic point groups include the inversion as well, that must be handled differently. Since the grain boundary space is defined in this article as involving only proper rotations, and applying the inversion to a single grain would make the misorientation an improper rotation, this would map a boundary to a point outside the space and not result in an observable symmetry. That said, applying the inversion to both grains preserves the misorientation but inverts the crystallographic directions aligned with the boundary plane normal \cite{patala2013symmetries}. With reference to the MBP parameterization, the effect of this on the grain boundary space is identical to left multiplying $g_b$ (or equivalently $g_1$ and $g_2$) by a binary rotation about any axis in the $x$-$y$ plane. Equation \ref{eq:gb_rotations_l} suggests that the $y$ axis is a particularly convenient choice, corresponding to a matrix $\mat{A}$ where the $(a, b, \gamma, \alpha, \beta)$th row contains a $(-1)^{a + b}$ in the $(a, b, -\gamma, \alpha, \beta)$th column and zeros elsewhere.

\subsection{Grain exchange symmetry}
\label{subsec:grain_exchange}

\begin{figure}
\center
\subfloat[]{%
	\label{subfig:initial_axes}{%
		\includegraphics[width=4.5cm]{%
			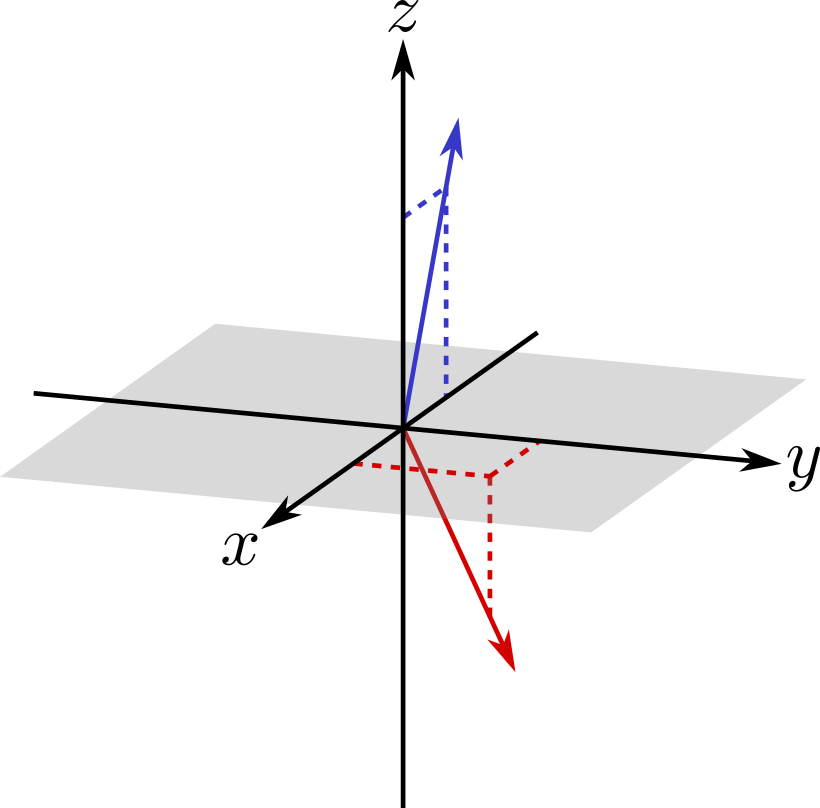}}} \quad
\subfloat[]{%
	\label{subfig:exchanged_axes}{%
		\includegraphics[width=4.5cm]{%
			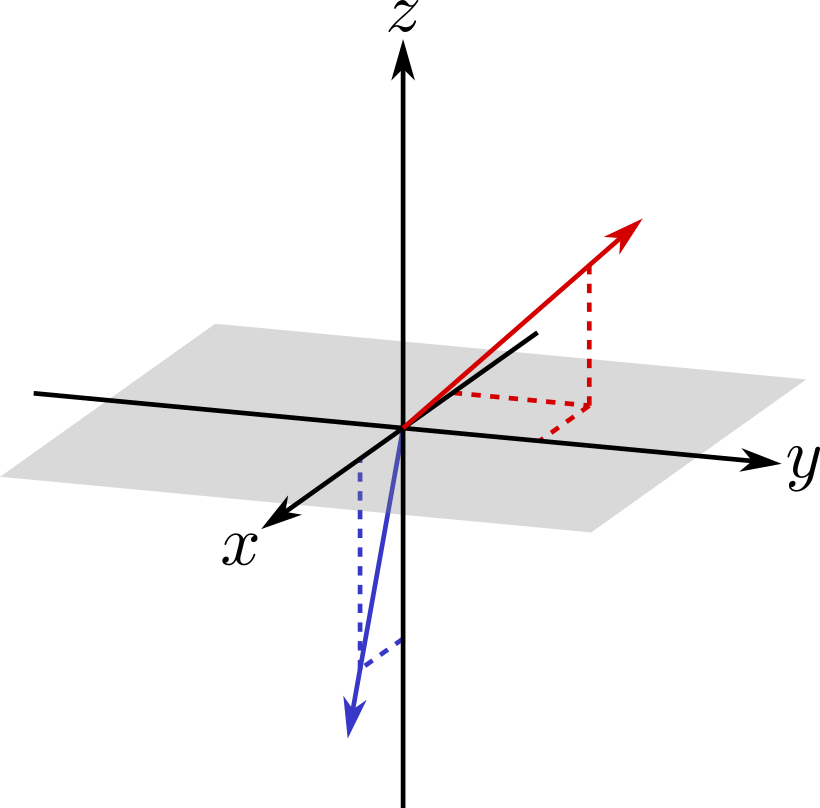}}}
\caption{\label{fig:grain_exchange}Two grain boundaries related by the grain exchange symmetry. (a) The red and blue axes are crystal directions aligned with the axes of rotation of grains one and two, respectively. (b) Rotating the entire configuration by $\pi$ about the $y$ axis and exchanging the grain labels gives a physically identical boundary. }
\end{figure}

The grain exchange symmetry follows from the observation that when considering homophase boundaries, the designation of one of the grains adjoining a grain boundary as grain one and the other as grain two is entirely arbitrary; any function on the grain boundary space should be invariant to this choice. The mathematical expression for this begins by parameterizing the grain rotation from the reference orientation by an axis and angle. Suppose that the shaded area in Fig.\ \ref{subfig:initial_axes} is the grain boundary plane, and the red and blue axes are crystal directions aligned with the rotations axes of grains one and two, respectively. Rotating the entire configuration by $\pi$ about the $y$ axis gives the physically identical boundary in Fig.\ \ref{subfig:exchanged_axes}, and further exchanging the grain labels realizes the grain exchange symmetry. This suggests that any function of the pair $(g_1, g_2)$ should have the same value at $(g_{\pi y} g_2, g_{\pi y} g_1)$, or
\begin{equation*}
\sum_{a b} \sum_{\alpha \beta \gamma} M^{a b}_{\alpha \beta \gamma}(g_1, g_2) c^{a b}_{\alpha \beta \gamma} = \sum_{a b} \sum_{\alpha \beta \gamma} M^{a b}_{\alpha \beta \gamma}(g_{\pi y} g_2, g_{\pi y} g_1) c^{a b}_{\alpha \beta \gamma}.
\end{equation*}
Expanding the basis functions with Eq.\ \ref{eq:numerical} and separating the constituent rotations with Eq.\ \ref{eq:su2_addition} gives
\begin{equation*}
\sum_{a b} \sum_{\alpha \beta \gamma} \Pi_{a b} U^{a}_{\alpha \gamma}(g^{-1}_1) U^{b}_{\beta -\gamma}(g^{-1}_2) c^{a b}_{\alpha \beta \gamma} = \sum_{a b} \sum_{\substack{\alpha \beta \gamma \\ \delta \epsilon}} \Pi_{a b} U^{a}_{\alpha \delta}(g^{-1}_2) U^{a}_{\delta \gamma}(g^{-1}_{\pi y}) U^{b}_{\beta \epsilon}(g^{-1}_1) U^{b}_{\epsilon -\gamma}(g^{-1}_{\pi y}) c^{a b}_{\alpha \beta \gamma}.
\end{equation*}
Since $U^{a}_{\alpha' \alpha}(g^{-1}_{\pi y}) = (-1)^{a - \alpha} \delta_{\alpha' -\alpha}$, substituting for the binary rotations and performing the summations over $\delta$ and $\epsilon$ yields
\begin{equation*}
\sum_{a b} \sum_{\alpha \beta \gamma} \Pi_{a b} U^{a}_{\alpha \gamma}(g^{-1}_1) U^{b}_{\beta -\gamma}(g^{-1}_2) c^{a b}_{\alpha \beta \gamma} = \sum_{a b} \sum_{\alpha \beta \gamma} \Pi_{a b} (-1)^{a + b} U^{a}_{\alpha -\gamma}(g^{-1}_2) U^{b}_{\beta \gamma}(g^{-1}_1) c^{a b}_{\alpha \beta \gamma}.
\end{equation*}
The dependence on the variables $g^{-1}_1$ and $g^{-1}_2$ can be handled by multiplying through by $U^{d*}_{\delta' \delta}(g^{-1}_1) U^{e*}_{\epsilon' -\delta}(g^{-1}_2)$ and using the orthogonality relation Eq.\ \ref{eq:su2_orthogonal}. Performing all of the remaining summations and relabelling the indices reduces this to
\begin{equation*}
c^{a b}_{\alpha \beta \gamma} = (-1)^{a + b} c^{b a}_{\beta \alpha \gamma}
\end{equation*}
for the condition to enforce the grain exchange symmetry. This corresponds to a transformation matrix $\mat{A}$ where the $(a, b, \gamma, \alpha, \beta)$th row contains a $(-1)^{a + b}$ in the $(b, a, \gamma, \beta, \alpha)$th column and zeros elsewhere.

\subsection{Null boundary symmetry}
\label{subsec:null_boundary}

As discussed in Section \ref{sec:grain_boundaries}, the usual procedure in the literature when considering the null boundary subspace (where the misorientation is the identity rotation) is to contract the subspace to a point and deform the surrounding space to maintain continuity. Instead, the procedure adopted here is to constrain the basis functions to be constant on the null boundary subspace; this achieves the equivalent result of requiring all boundaries on the subspace to have identical properties without some arbitrary choice of deformation.

Since the proposed constraint explicitly involves the misorientation, the MBP parameterization of the basis functions on the grain boundary space is adopted. A function on the grain boundary space restricted to the null misorientation can be written as the expansion
\begin{equation*}
f(g_b) = \sum_{a b} \sum_{\alpha \beta \gamma} M^{a b}_{\alpha \beta \gamma}(g_b, g_b) c^{a b}_{\alpha \beta \gamma}.
\end{equation*}
By hypothesis, $f(g_b)$ is equal to a constant $f_0$. Since the spherical harmonics form a complete orthonormal basis for functions on $S^2$, the proposed constraint can be written as
\begin{equation*}
\int^{2 \pi}_{0} \int^{\pi}_{0} f(g_b) Y^{\epsilon' *}_{e'}(\omega_b, \phi_b) \sin \omega_b \diff \omega_b \diff \phi_b = 2 \sqrt{\pi} f_0 \delta_{e' 0} \delta_{\epsilon' 0}.
\end{equation*}
Substituting the expansion for $f(g_b)$ and exchanging the order of integration and summation gives
\begin{equation}
\sum_{a b} \sum_{\alpha \beta \gamma} \bigg[ \int^{2 \pi}_{0} \int^{\pi}_{0} M^{a b}_{\alpha \beta \gamma}(g_b, g_b) Y^{\epsilon' *}_{e'}(\omega_b, \phi_b) \sin \omega_b \diff \omega_b \diff \phi_b \bigg] c^{a b}_{\alpha \beta \gamma} = 2 \sqrt{\pi} f_0 \delta_{e' 0} \delta_{\epsilon' 0}.
\label{eq:null_constraint}
\end{equation}
Provided the quantity in brackets can be found, Eq.\ \ref{eq:null_constraint} gives a linear system of equations that can be solved for the space of coefficients that satisfy the null boundary symmetry.

Consider the basis functions on the grain boundary space restricted to the null misorientation. Substituting Eq.\ \ref{eq:define_z} for the hyperspherical harmonics in Eq.\ \ref{eq:MBP_reduced}, using the property of the generalized characters that $\chi^{l}_{\lambda}(0) = \Pi^2_{l} \delta_{\lambda 0}$ \cite{1988varshalovich}, and performing the summations over $f$ and $\phi$ gives
\begin{equation*}
M^{a b}_{\alpha \beta \gamma}(g_b, g_b) = \sum_{e} \sum_{\delta \epsilon} \frac{2 \sqrt{2} \Pi_{a b}}{\sqrt{\pi} \Pi_{e}} (-i)^{\epsilon} C^{e -\epsilon}_{a \alpha b \delta} C^{b \delta}_{0 0 b \beta} C^{e 0}_{a \gamma b -\gamma} Y^{0}_{0}(\omega_m, \phi_m) Y^{\epsilon}_{e}(\omega_b, \phi_b).
\end{equation*}
This can be further simplified by evaluating $Y^{0}_{0}(\omega_m, \phi_m)$, using one of the properties of the Clebsh--Gordan coefficients, and performing the summation over $\delta$ to find 
\begin{equation*}
M^{a b}_{\alpha \beta \gamma}(g_b, g_b) = \sum_{e \epsilon} \frac{\sqrt{2} \Pi_{a b}}{\pi \Pi_{e}} (-i)^{\epsilon} C^{e -\epsilon}_{a \alpha b \beta} C^{e 0}_{a \gamma b -\gamma} Y^{\epsilon}_{e}(\omega_b, \phi_b).
\end{equation*}
Inserting this into Eq.\ \ref{eq:null_constraint}, exchanging the order of integration and summation, using the orthonormality of the spherical harmonics, multiplying though by several constants, and finally relabelling the indices gives
\begin{equation*}
\sum_{a b} \sum_{\alpha \beta \gamma} \bigg[ \Pi_{a b} C^{e -\epsilon}_{a \alpha b \beta} C^{e 0}_{a \gamma b -\gamma} \bigg] c^{a b}_{\alpha \beta \gamma} = \sqrt{2 \pi^3} f_0 \delta_{e 0} \delta_{\epsilon 0}
\end{equation*}
for the system of equations to be solved. If $f_0 = 0$, as for the energy of a homophase grain boundary, then the subspace of coefficients consistent with the null boundary symmetry is equivalent to the nullspace of the matrix in brackets. If this matrix is indicated by $\mat{B}$, then
\begin{equation*}
\mat{X}_{i + 1} = \mat{X}_{i} \mathrm{N}(\mat{B} \mat{X}_{i})
\end{equation*}
is suggested as a more efficient alternative to Eq.\ \ref{eq:add_symmetry} for the null boundary symmetry; this retains only the components of the basis for the nullspace of $\mat{B}$ fully within the span of $\mat{X}_{i}$.

\subsection{Example: Octahedral point group}
\label{subsec:octahedral}

\begin{figure}
    \centering
    \includegraphics[width=9cm]{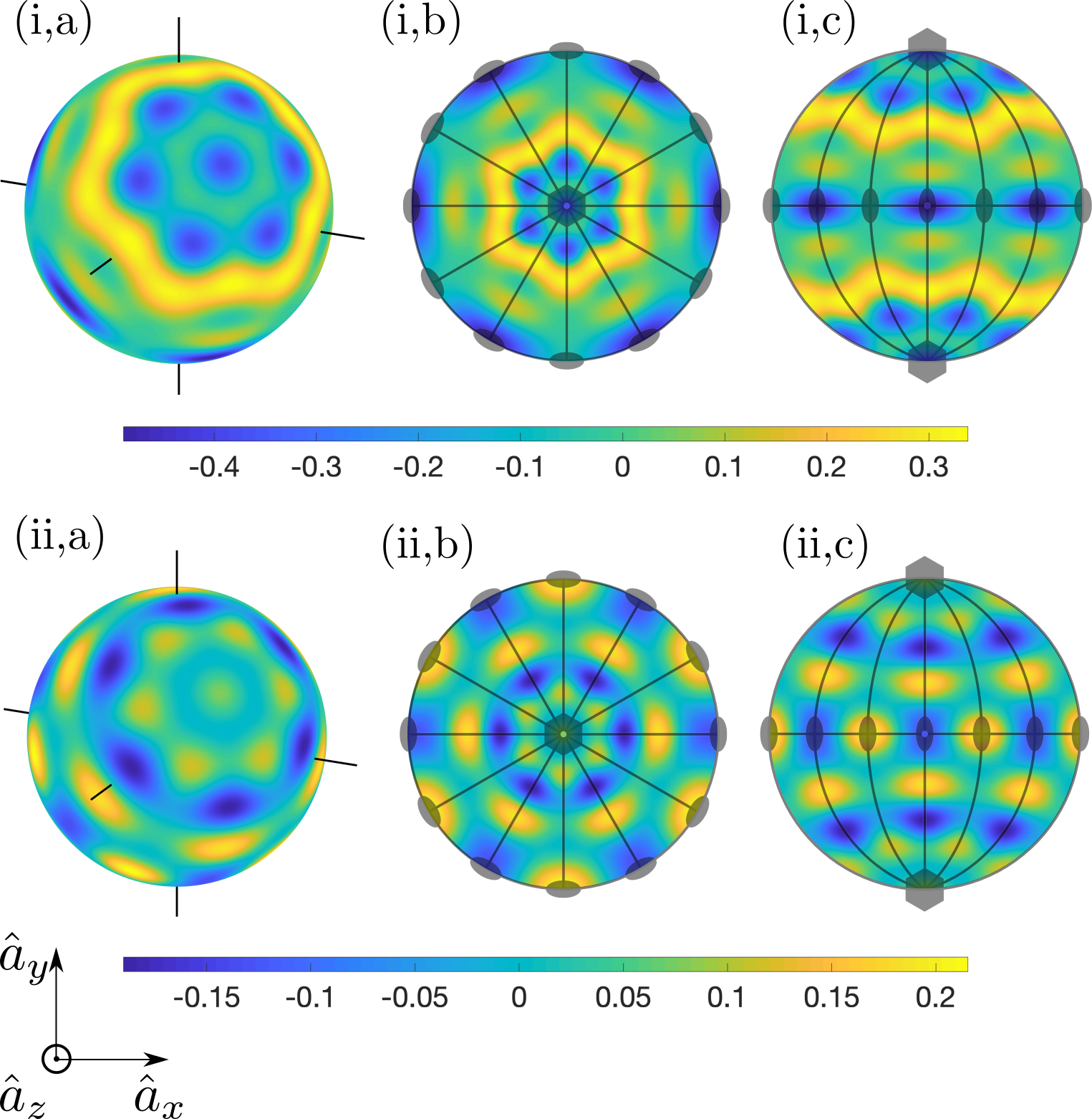}
    \caption{Basis functions plotted in the boundary-plane inclination space for the $\Sigma 3$ misorientation. To illustrate the $D_{6h}$ symmetry elements, the basis functions are projected along the $\hat{a}_z = [111]$ and $\hat{a}_z = [2 \bar{1} \bar{1}]$ axes in (b) and (c), respectively.}
    \label{fig:d6h_symm}
\end{figure}

As evidence that the symmetrization procedure performs as intended, this section provides examples of basis functions for homophase grain boundaries in a cubic material with octahedral point group symmetry. Specifically, the symmetries exhibited by the boundary-plane inclinations for fixed misorientations are illustrated. On the basis of bicrystallography, additional symmetries arise in the space of boundary-plane inclinations when the crystal and grain-exchange symmetries are considered \cite{patala2013symmetries}. For example, in crystals with the octahedral point group symmetry $\left( O_h \right)$, if the misorientation is $\left( \left[ 1 \, 1 \, 1 \right], 60^{\circ} \right)$ (the $\Sigma 3$ misorientation) then the boundary-plane inclinations exhibit $D_{6h}$ point group symmetry \cite{banadaki2016simple} with $6$-fold and $2$-fold axes aligned with the $[111]$ and $[2 \bar{1} \bar{1}]$ directions, respectively. Figure \ref{fig:d6h_symm} shows two symmetrized basis functions for the $\Sigma 3$ misorientation (computed for $N_{\text{max}} = \max(a+b) = 8$); since the remaining variables specify the boundary-plane normal vector $\hat{n}$, the basis-functions are plotted on the unit sphere. Figures \ref{fig:d6h_symm}(b) and \ref{fig:d6h_symm}(c) show the functions projected along the $[111]$ and the $[2 \bar{1} \bar{1}]$ axes, clearly revealing the $D_{6h}$ symmetry exhibited by the boundary-plane inclinations for this misorientation.

\begin{figure}
    \centering
    \includegraphics[width=9cm]{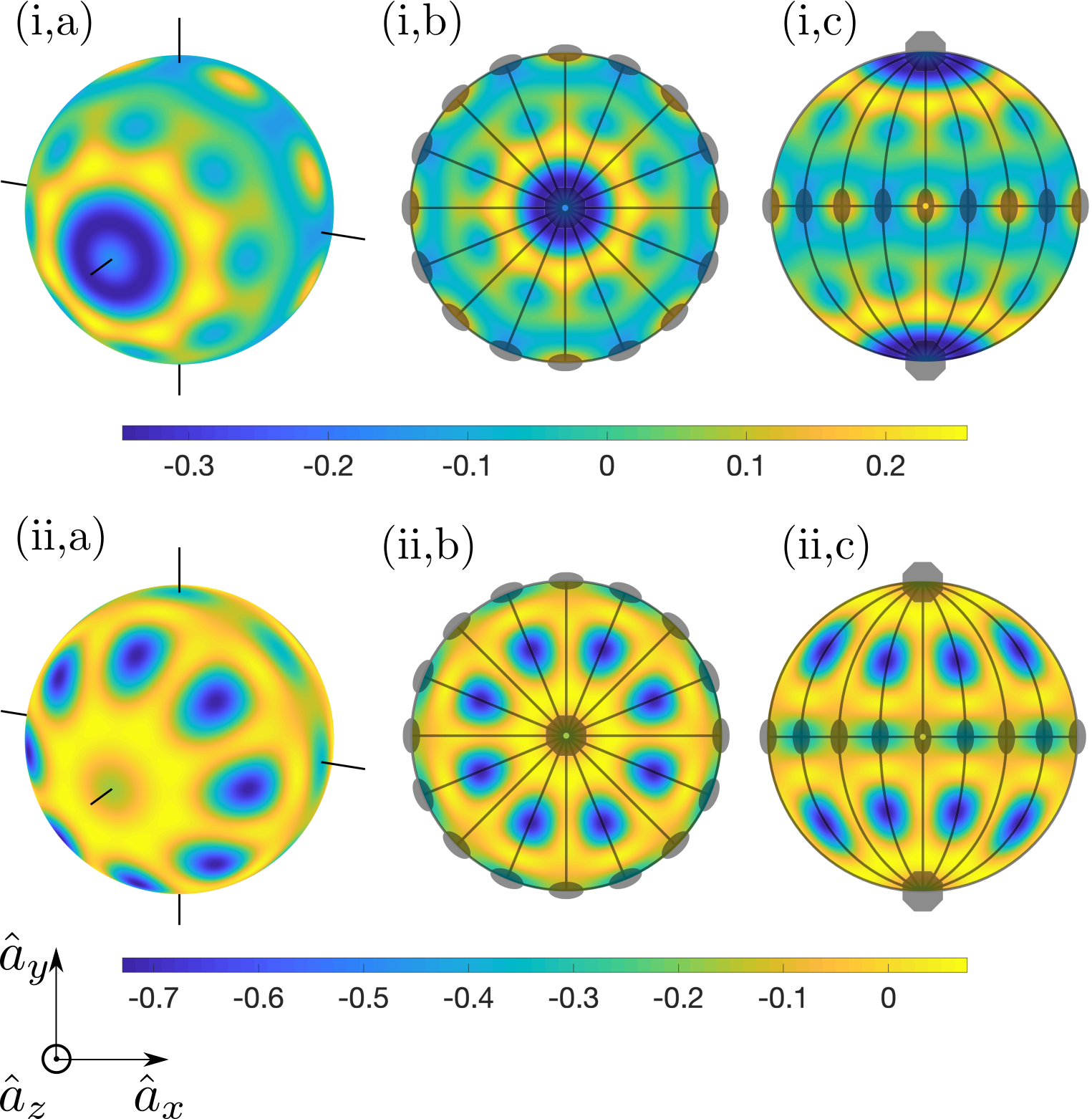}
    \caption{Basis functions plotted in the boundary-plane inclination space for the $\left( \left[ 1 \, 0 \, 0 \right]; 45^{\circ} \right)$ misorientation. To illustrate the $D_{8h}$ symmetry elements, the basis functions are projected along the $\hat{a}_z = [100]$ and $\hat{a}_z = [0 \, \cos (\pi/8) \, \sin (\pi/8)]$ axes in (b) and (c), respectively.}
    \label{fig:d8h_symm}
\end{figure}

Another example is the $D_{8h}$ symmetry exhibited by the boundary-plane inclinations for the $\left( \left[ 1 \, 0 \, 0 \right], 45^{\circ} \right)$ misorientation with 8-fold and the 2-fold symmetry axes aligned with the $[100]$ and $[0 \, \cos (\pi/8) \, \sin(\pi/8)]$ directions, respectively. Figure \ref{fig:d8h_symm} shows the same basis functions as in Figure \ref{fig:d6h_symm} but for the $\left( \left[ 1,0,0 \right], 45^{\circ} \right)$ misorientation, and the projections along the 8-fold and the 2-fold symmetry axes illustrate the $D_{8h}$ symmetry. These two examples validate the symmetrization framework developed for the grain boundary basis functions.

\begin{figure}
    \centering
    \includegraphics[width=9cm]{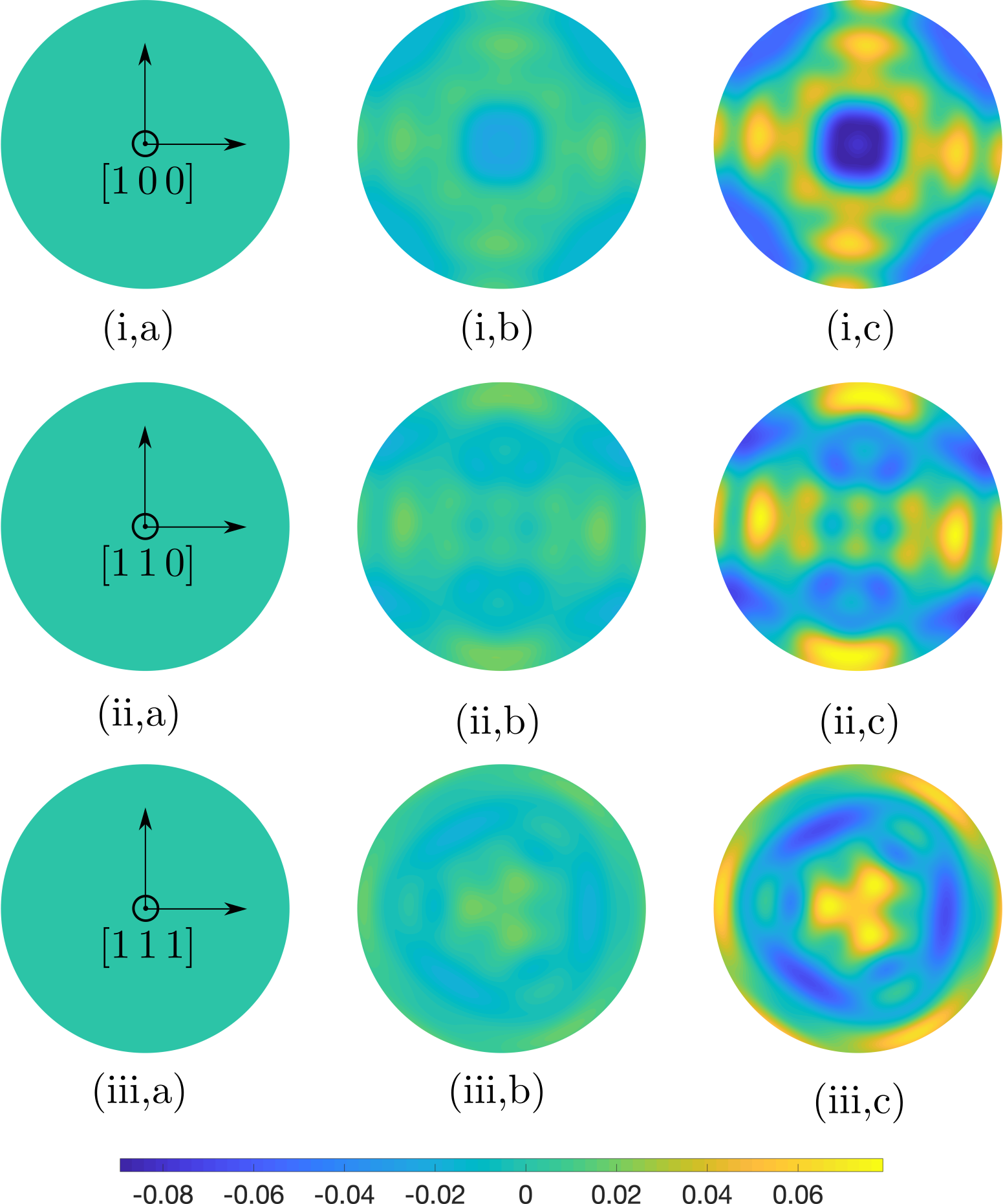}
    \caption{Example symmetrized basis function on the boundary-plane inclination space projected along the misorientation axes (i) $[100]$, (ii) $[110]$ and (iii) $[111]$ for the misorientation angles (a) $\omega = 0^{\circ}$, (b) $\omega = 5^{\circ}$ and (c) $\omega = 10^{\circ}$. The basis function is constructed to satisfy the null-boundary constraint, that is, to approach zero as the misorientation angle approaches zero; this is visible from the plots in (a) for all three misorientation axes.}
    \label{fig:null_gb}
\end{figure}

Finally, we provide plots showing the consistency of the symmetrized basis functions around the null-boundary singularity. Figure \ref{fig:null_gb} shows an example basis function for the misorientation axes (i) $[100]$, (ii) $[110]$, (iii) $[111]$ and misorientation angles (a) $0^{\circ}$, (b) $5^{\circ}$ and (c) $10^{\circ}$ by means of stereographic projections along the misorientation axis. Since the boundary-plane space contains an inversion-center symmetry for the $O_h$ point group, it suffices to project just the top-half of the sphere. The figure indicates that, irrespective of the boundary-plane orientation, the value of the basis function approaches zero as the misorientation angle approaches zero. While the figure only shows a single basis function along three misorientation axes, we have verified that all the symmetrized basis functions go to zero when the identity misorientation is approached along any axis of rotation. That is, the question of the null-boundary singularity is resolved by the symmetrized basis functions.

\section{Conclusions}
\label{sec:conclusions}

A polycrystalline material's texture can have a profound effect on the anisotropy of the material's mechanical, thermal, electronic, and magnetic properties. Moreover, the texture influences and is influenced by the microstructure evolution that occurs during thermomechanical processing, and as such can be used to infer a material's processing history. The basis functions on the space of rotations \cite{1982bunge,2009mason_b} are essential to these applications, constituting the mathematical language connecting the various parts of the discipline. They allow the material's texture to be reconstructed from diffraction measurements, provide equations to predict the anisotropy of the material's properties, and suggest a standard format for texture information to be made available to the community.

The basis functions on the grain boundary space developed in Sections \ref{sec:grain_boundaries} and \ref{sec:symmetry} are envisioned as providing analogous mathematical foundations for the emerging discipline of grain boundary analysis. The extent of the grain boundary space makes systematic study of grain boundary properties difficult, with the result that the majority of studies to date have been concerned with particular boundaries or families of boundaries believed to be most relevant to material properties. Real materials are of course subject to no such limitations, and contain boundaries with a wide variety of characters and generally unknown properties.

The authors are of the opinion that the most expedient way to close this knowledge gap lies in the acquisition of sufficient data regarding the distribution and properties of grain boundaries in real materials. While the relevant experimental techniques are arguably already in place, the ability to quantitatively connect the grain boundary character distribution to, e.g., susceptibility to intergranular fracture, is not yet conclusively established. The authors hope that the results presented here facilitate such a development in the near future.

\appendix

\section{Rotation conventions}
\label{app:convention}

Let a set of orthonormal basis vectors $\vec{e}_i$ be specified in a fixed laboratory frame. A vector $\vec{r}$ in $\mathbb{R}^3$ can be described as the product of a row vector $\langle \vec{e} | = \begin{bmatrix}[0.75] \vec{e}_x & \vec{e}_y & \vec{e}_z \end{bmatrix}$ of the basis vectors and a column vector $| r \rangle = \begin{bmatrix}[0.75] x & y & z \end{bmatrix}^{T}$ of the spatial coordinates:
\begin{equation*}
\vec{r} = \langle \vec{e} | r \rangle.
\end{equation*}
If $\vec{r}$ is rotated in the laboratory frame by some rotation $g_1$ before a second rotation $g_2$, the resulting vector $\vec{r}'$ is given by the matrix product\begin{equation*}
\vec{r}' = \langle \vec{e} | \mat{R}(g_2) \mat{R}(g_1) | r \rangle
\end{equation*}
where $\mat{R}(g)$ is a rotation matrix. When the rotation is parameterized by the rotation angle $\omega$ and a unit vector $\vec{n}$ along the rotation axis with components measured in the laboratory frame, $\mat{R}(g)$ can be evaluated as \cite{1986altmann}
\begin{align*}
\mat{Z} &= \begin{bmatrix}[0.75] 0 & -n_z & n_y \\ n_z & 0 & -n_x \\ -n_y & n_x & 0 \end{bmatrix} \\
\mat{R}(\omega, \vec{n}) &= \mat{I} + \sin \omega \mat{Z} + (1 - \cos \omega) \mat{Z}^{2}
\end{align*}
where $\mat{I}$ is the identity matrix. Functions on the unit sphere in $\mathbb{R}^3$ transform in the same way, but the basis is constructed of the spherical harmonics $Y^{\lambda''}_{l}(\theta, \phi)$, the transformation matrices $U^{l}_{\lambda'' \lambda'}(g)$ are the irreducible representations of \so{3}, and the coordinates are the coefficients of the expansion $c_{l \lambda}$:
\begin{equation*}
f'(\theta, \phi) = \sum_{l} \sum_{\lambda'' \lambda' \lambda} Y^{\lambda''}_{l}(\theta, \phi) U^{l}_{\lambda'' \lambda'}(g_2) U^{l}_{\lambda' \lambda}(g_1) c_{l \lambda}.
\end{equation*}
Functions on other spaces transform in the same way by construction, meaning that this convention is used to identify collections of terms that constitute the function bases and transformation matrices.

\section{Standard relations}
\label{app:standard}

This section provides several standard relations for the benefit of the reader. Summations are performed over all values for which the summand is nonzero, and $\Pi_{a \dots b}$ stands for the product $[(2 a + 1) \dots (2 b + 1)]^{1/2}$. The hyperspherical harmonics are defined as \cite{1967domokos,1976kalnins,2009mason_a}
\begin{align}
Z^{n}_{l \lambda}(g) &= (-i)^{l} \frac{2^{l + 1 / 2} l!}{2 \pi} \left[ (2 l + 1) \frac{(l - \lambda)!}{(l + \lambda)!} \frac{(n + 1) (n - l)!}{(n + l + 1)!} \right]^{1/2} \sin^{l}(\omega / 2) C^{l + 1}_{n - l}[ \cos(\omega / 2) ] \nonumber \\
&\quad P^{\lambda}_{l}(\cos \theta) e^{i \lambda \phi} \nonumber \\
&= \sqrt{2 / \pi} (-i)^{l} \chi^{n / 2}_{l}(\omega) Y^{\lambda}_{l}(\theta, \phi)
\label{eq:define_z}
\end{align}
where the rotation $g$ is parameterized by the rotation angle $\omega$ and the polar and azimuthal angles $\theta$ and $\phi$ of the rotation axis. $C^{l + 1}_{n - l}[ \cos(\omega / 2) ]$ is a Gegenbauer polynomial, $P^{\lambda}_{l}(\cos \theta)$ is an associated Legendre polynomial with the Condon--Shortley phase convention, $\chi^{n / 2}_{l}(\omega)$ is a generalized character of the irreducible representations (irreps) of \so{3} \cite{1988varshalovich}, and $Y^{\lambda}_{l}(\theta, \phi)$ is a spherical harmonic. The hyperspherical harmonics are orthonormal with respect to the differential volume element given by Eq.\ \ref{eq:su2_volume}, and are related to the irreps of \su{2} by \cite{2009mason_a}
\begin{align}
Z^{2 l}_{m \mu}(g) &= \sum_{\lambda' \lambda} \frac{\Pi_m}{\sqrt{2} \pi} C^{l \lambda}_{l \lambda' m \mu} U^{l}_{\lambda' \lambda}(g) \label{eq:z_to_su2} \\
U^{l}_{\lambda' \lambda}(g) &= \sum_{m \mu} \sqrt{2} \pi \frac{\Pi_{m}}{\Pi^2_{l}} C^{l \lambda}_{l \lambda' m \mu} Z^{2 l}_{m \mu}(g)
\label{eq:su2_to_z}
\end{align}
where $C^{l \lambda}_{l \lambda' m \mu}$ is a Clebsch--Gordan coefficient. Inverting the rotation $g$ by changing the sign of the rotation angle gives the following transformations of the hyperspherical harmonics (with Eq.\ \ref{eq:define_z}) and the irreps of \su{2} (with Eqs.\ \ref{eq:su2_to_z} and \ref{eq:invert_z}):
\begin{align}
Z^{n}_{l \lambda}(g^{-1}) &= (-1)^{l} Z^{n}_{l \lambda}(g) \label{eq:invert_z} \\
U^{l}_{\lambda' \lambda}(g^{-1}) &= (-1)^{\lambda' - \lambda} U^{l}_{-\lambda -\lambda'}(g). \label{eq:invert_su2}
\end{align}
Manipulating the irreps of \su{2} can be more convenient than the hyperspherical harmonics due to the relative simplicity of the addition theorem and Clebsh--Gordan expansion of the former \cite{1988varshalovich}:
\begin{align}
\sum_{\mu} U^{l}_{\lambda' \mu}(g_2) U^{l}_{\mu \lambda}(g_1) &= U^{l}_{\lambda' \lambda}(g_2 g_1) \label{eq:su2_addition} \\
U^{l}_{\lambda' \lambda}(g) U^{m}_{\mu' \mu}(g) &= \sum_{n \nu' \nu} C^{n \nu'}_{l \lambda' m \mu'} U^{n}_{\nu' \nu}(g) C^{n \nu}_{l \lambda m \mu}. \label{eq:clebsh_gordan}
\end{align}
Notice that the composition of rotations is from right to left, following the convention in Appendix \ref{app:convention}. The relationship of the spherical harmonics to the irreps of \so{3} gives
\begin{equation}
Y^{\lambda}_{l}(\theta, \phi) = (-1)^{\lambda} \frac{\Pi_{l}}{2 \sqrt{\pi}} d^{l}_{-\lambda 0}(\theta) e^{i \lambda \phi}
\label{eq:spherical_harmonic}
\end{equation}
as an alternate expression for the spherical harmonics, where $d^{l}_{\lambda' \lambda}(\theta)$ is known as a Wigner little $d$-function \cite{1988varshalovich}. Finally, the sum
\begin{equation}
\sum_{c} C^{c 0}_{a \alpha' b -\alpha'} C^{c 0}_{a \alpha b -\alpha} = \delta_{\alpha' \alpha}
\label{eq:clebsh_gordan_sum}
\end{equation}
is a special case of one of the standard sums involving a product of two Clebsch--Gordan coefficients \cite{1988varshalovich}.

\section{Volume of the rotation group}
\label{app:rotation_volume}

Let $\mathbb{Q}$ be the group of normalized quaternions on the unit sphere in four dimensions $S^3$. Since the volume of the sphere is $2 \pi^2$ in the Euclidean metric, and $\mathbb{Q}$ is isomorphic to \su{2}, this suggests that the volumes of $\mathbb{Q}$ and \su{2} should be $2 \pi^2$ as well. This differs from the standard value of $16 \pi^2$ by a factor of $1 / 8$, and requires that several of the standard orthogonality conditions be adjusted to follow this convention.

For specificity, let a rotation $g$ be parameterized by the rotation angle $\omega$ and the polar and azimuthal angles $\theta$ and $\phi$ of the rotation axis. With this parameterization, integration over \su{2} is performed with the differential volume element
\begin{equation}
\diff \Omega = \frac{1}{2} \sin^2( \omega / 2 ) \sin \theta \diff \omega \diff \theta \diff \phi
\label{eq:su2_volume}
\end{equation}
over the intervals $[0, 2 \pi]$, $[0, \pi]$, and $[0, 2\pi]$, respectively. This is consistent with the volume of \su{2} as specified above:
\begin{equation}
\int_{0}^{2 \pi} \int_{0}^{\pi} \int_{0}^{2 \pi} \frac{1}{2} \sin^2( \omega / 2 ) \sin \theta \diff \omega \diff \theta \diff \phi = 2 \pi^2. \nonumber
\end{equation}
The hyperspherical harmonics as defined in Eq.\ \ref{eq:define_z} are normalized with respect to the same integration over \su{2}:
\begin{equation}
\int Z^{n'}_{l' \lambda'}( g ) Z^{n*}_{l \lambda}( g ) \diff \Omega = \delta_{n' n} \delta_{l' l} \delta_{\lambda' \lambda}. \label{eq:z_orthogonal}
\end{equation}
The normalization of the hyperspherical harmonics and their relationship to the matrix elements of the irreps of \su{2}, as given by Eq.\ \ref{eq:su2_to_z}, defines the orthogonality relation for the matrix elements of the irreps of \su{2}:
\begin{align}
\int U^{a}_{\alpha' \alpha}( g ) U^{b*}_{\beta' \beta}( g ) \diff \Omega &= \sum_{\mathclap{c \gamma d \delta}} 2 \pi^2 \frac{\Pi_{c d}}{\Pi^2_{a b}} C^{a \alpha}_{a \alpha' c \gamma} C^{b \beta}_{b \beta' d \delta} \int Z^{2 a}_{c \gamma}( g ) Z^{2 b*}_{d \delta}( g ) \diff \Omega \nonumber \\
&= \sum_{c \gamma} 2 \pi^2 \frac{\Pi^2_{c}}{\Pi^4_{a}} C^{a \alpha}_{a \alpha' c \gamma} C^{a \beta}_{a \beta' c \gamma} \delta_{a b} \nonumber \\
&= \frac{2 \pi^2}{\Pi^2_{a}} \delta_{a b} \delta_{\alpha' \beta'} \delta_{\alpha \beta}.
\label{eq:su2_orthogonal}
\end{align}
Observe that this orthogonality relation and the volume of \su{2} as defined in this article differ from the standard orthogonality relation and the standard volume of \su{2} \cite{1988varshalovich} by the same factor of $1 / 8$, and that Eq.\ \ref{eq:su2_orthogonal} is independent of the parameterization of \su{2}.

Let $\diff \Omega_1 \diff \Omega_2$ indicate the differential volume element on \so{4}. Expanding the inner product of matrix elements of the irreps of \so{4} as defined in Eq.\ \ref{eq:so4_matrix} gives
\begin{align}
\iint R^{a' b'}_{c' \gamma' d' \delta'}( g_1, g_2 ) R^{a b*}_{c \gamma d \delta}( g_1, g_2 ) \diff \Omega_1 \diff \Omega_2 = \sum_{\mathclap{\substack{\alpha \beta \mu \nu \\ \epsilon \phi \psi \eta}}} & C^{c' \gamma'}_{a' \alpha b' \beta} C^{d' \delta'}_{a' \mu b' \nu} C^{c \gamma}_{a \epsilon b \phi} C^{d \delta}_{a \psi b \eta} \nonumber \\
& \int U^{a'}_{\alpha \mu}( g^{-1}_2 ) U^{a*}_{\epsilon \psi}( g^{-1}_2 ) \diff \Omega_1 \int U^{b'}_{\beta \nu}( g_1 ) U^{b*}_{\phi \eta}( g_1 ) \diff \Omega_2 \nonumber
\end{align}
Integrating the matrix elements of the irreps of \su{2} with Eq.\ \ref{eq:su2_orthogonal}, summing over $\epsilon$, $\phi$, $\psi$ and $\eta$, and then over $\alpha$, $\beta$, $\mu$ and $\nu$, gives the orthogonality relation for the matrix elements of the irreps of \so{4} as
\begin{equation}
\iint R^{a' b'}_{c' \gamma' d' \delta'}( g_1, g_2 ) R^{a b*}_{c \gamma d \delta}( g_1, g_2 ) \diff \Omega_1 \diff \Omega_2 = \frac{4 \pi^4}{\Pi^2_{a b}} \delta_{a' a} \delta_{b' b} \delta_{c' c} \delta_{\gamma' \gamma} \delta_{d' d}  \delta_{\delta' \delta}.
\label{eq:so4_orthogonal}
\end{equation}

\section{NNT metric}
\label{app:NNT_metric}

Following the reasoning in Sec.\ \ref{subsec:NNT_param}, the rotations $g_1 = g_z g_{n1}$ and $g_2 = g_z g_t g_{n2}$ correspond to points on $S^3 \times S^3$ embedded in $\mathbb{R}^8$ with the following Cartesian coordinates:
\begin{align*}
w_1 &= \cos( \omega_{n1} / 2 ) \cos( \omega_z / 2 ) & w_2 &= \cos( \omega_{n2} / 2 ) \cos( \omega_t / 2 + \omega_z / 2 ) \\
x_1 &= \sin( \omega_{n1} / 2 ) \cos( \phi_{n1} + \omega_z / 2 ) & x_2 &= \sin( \omega_{n2} / 2 ) \cos( \phi_{n2} + \omega_t / 2 + \omega_z / 2 ) \\
y_1 &= \sin( \omega_{n1} / 2 ) \sin( \phi_{n1} + \omega_z / 2 ) & y_2 &= \sin( \omega_{n2} / 2 ) \sin( \phi_{n2} + \omega_t / 2 + \omega_z / 2 ) \\
z_1 &= \cos( \omega_{n1} / 2 ) \sin( \omega_z / 2 ) & z_2 &= \cos( \omega_{n2} / 2 ) \sin( \omega_t / 2 + \omega_z / 2 ).
\end{align*}
This is found by writing, e.g., the rotations $g_z$, $g_t$, and $g_{n2}$ as unit quaternions, performing the quaternion multiplication, and interpreting the result $g_2$ as a point on $S^3$. Transforming the Euclidean metric on $\mathbb{R}^8$ with the Jacobian matrix of the above coordinate change gives the following nonzero components for the upper triangular part of the metric tensor on $\su{2} \times \su{2}$:
\begin{align*}
g_{11} &= 1 / 4 & g_{44} &= \sin^2( \omega_{n2} / 2 ) & g_{56} &= 1 / 4 \\
g_{22} &= \sin^2 ( \omega_{n1} / 2 ) & g_{45} &= ( 1 - \cos \omega_{n2} ) / 4 & g_{66} &= 1 / 2 \\
g_{26} &= ( 1 - \cos \omega_{n1} ) / 4 & g_{46} &= ( 1 - \cos \omega_{n2} ) / 4 \\
g_{33} &= 1 / 4 & g_{55} &= 1 / 4.
\end{align*}
The indices correspond to $\omega_{n1}$, $\phi_{n1}$, $\omega_{n2}$, $\phi_{n2}$, $\omega_t$, and $\omega_z$ in increasing order. Taking the square root of the determinant of the metric tensor shows that the differential volume element on the space is
\begin{equation*}
\diff \Omega = \frac{1}{64} \abs{ \sin \omega_{n1} } \abs{ \sin \omega_{n2} } \diff \omega_{n1} \diff \phi_{n1} \diff \omega_{n2} \diff \phi_{n2} \diff \omega_t \diff \omega_z.
\end{equation*}
Integrating the volume element over the intervals $[0, 2 \pi]$, $[0, 2 \pi]$, $[0, 2 \pi]$, $[0, 2 \pi]$, $[0, 2 \pi]$, and $[0, 2 \pi]$ shows that the volume of the space is $4 \pi^4$, as expected.

\section{MBP metric}
\label{app:MBP_metric}

Following the reasoning in Sec.\ \ref{subsec:MBP_param}, the rotations $g_1 = g_z g_b$ and $g_2 = g_z g_b g_m$ correspond to points on $S^3 \times S^3$ embedded in $\mathbb{R}^8$ with the following Cartesian coordinates, where $\cs$ indicates a cosine and $\sn$ indicates a sine:
\begin{align*}
w_1 &= \cs( \omega_b / 2 ) \cs( \omega_z / 2 ) \\
x_1 &= \sn( \omega_b / 2 ) \cs( \phi_b + \omega_z / 2 ) \\
y_1 &= \sn( \omega_b / 2 ) \sn( \phi_b + \omega_z / 2 ) \\
z_1 &= \cs( \omega_b / 2 ) \sn( \omega_z / 2 ) \\
w_2 &= \cs( \omega_m  / 2 ) \cs( \omega_b / 2 ) \cs( \omega_z / 2 ) - \sn( \omega_m / 2 ) [ \cs \theta_m \cs( \omega_b / 2 ) \sn( \omega_z / 2 ) + \sn \theta_m \sn( \omega_b / 2 ) \cs( \phi_m - \phi_b - \omega_z / 2 ) ] \\
x_2 &= \cs( \omega_m / 2 ) \sn( \omega_b / 2 ) \cs( \phi_b + \omega_z / 2 ) \\
&\quad + \sn( \omega_m / 2 ) [ \sn \theta_m \cs( \omega_b / 2 ) \cs( \phi_m + \omega_z / 2 ) + \cs \theta_m \sn( \omega_b / 2 ) \sn( \phi_b + \omega_z / 2 ) ] \\
y_2 &= \cs( \omega_m / 2 ) \sn( \omega_b / 2 ) \sn( \phi_b + \omega_z / 2 ) \\
&\quad + \sn( \omega_m / 2 ) [ \sn \theta_m \cs( \omega_b / 2 ) \sn( \phi_m + \omega_z / 2 ) - \cs \theta_m \sn( \omega_b / 2 ) \cs( \phi_b + \omega_z / 2 ) ] \\
z_2 &= \cs( \omega_m / 2 ) \cs( \omega_b / 2 ) \sn( \omega_z / 2 ) + \sn( \omega_m / 2 ) [ \cs \theta_m \cs( \omega_b / 2 ) \cs( \omega_z / 2 ) + \sn \theta_m \sn( \omega_b / 2 ) \sn( \phi_m - \phi_b - \omega_z / 2 ) ].
\end{align*}
This is found by writing, e.g., the rotations $g_z$, $g_b$, and $g_m$ as unit quaternions, performing the quaternion multiplication, and interpreting the result $g_2$ as a point on $S^3$. Transforming the Euclidean metric on $\mathbb{R}^8$ with the Jacobian matrix of the above coordinate change gives the following nonzero components for the upper triangular part of the metric tensor on $\su{2} \times \su{2}$:
\begin{align*}
g_{11} &= 1 / 4 \\
g_{14} &= \sn \theta_m \cs(\phi_m - \phi_b) / 4 \\
g_{15} &= [\cs \theta_m (\cs \omega_b - 1) + \sn \theta_m \sn \omega_b \sn(\phi_m - \phi_b)] / 4 \\
g_{16} &= [\cs \theta_m \cs \omega_b + \sn \theta_m \sn \omega_b \sn(\phi_m - \phi_b)] / 4 \\
g_{22} &= \sn^2(\omega_m / 2) \\
g_{24} &= \sn(\omega_m / 2) [\cs(\omega_m / 2) \cs \theta_m \cs(\phi_m - \phi_b) - \sn(\omega_m / 2) \sn(\phi_m - \phi_b)] / 2 \\
g_{25} &= \sn(\omega_m / 2) \sn(\omega_b / 2) [ \cs(\omega_m / 2) \cs \theta_m \cs (\omega_b / 2) \sn(\phi_m - \phi _ b) + \cs(\omega_m / 2) \sn \theta_m \sn(\omega_b / 2) \\
& \phantom{= \sn(\omega_m / 2) \sn(\omega_b / 2) [} + \sn(\omega_m / 2) \cs(\omega_b / 2) \cs(\phi_m - \phi_b)] \\
g_{26} &= \sn(\omega_m / 2) [\cs(\omega_m / 2) \cs \theta_m \sn \omega_b \sn(\phi_m - \phi_b) - \cs(\omega_m / 2) \sn \theta_m \cs \omega_b + \sn(\omega_m / 2) \sn \omega_b \cs(\phi_m - \phi_b)] / 2 \\
g_{33} &= \sn^2(\omega_m / 2) \sn^2 \theta_m \\
g_{34} &= -\sn(\omega_m / 2) \sn \theta_m [\cs(\omega_m / 2) \sn(\phi_m - \phi_b) + \sn(\omega_m / 2) \cs \theta_m \cs(\phi_m - \phi_b)] / 2 \\
g_{35} &= -\sn(\omega_m / 2) \sn \theta_m \sn(\omega_b / 2) [ \sn(\omega_m / 2) \cs \theta_m \cs(\omega_b / 2) \sn(\phi_m - \phi_b) + \sn(\omega_m / 2) \sn \theta_m \sn(\omega_b / 2) \\
& \phantom{= -\sn(\omega_m / 2) \sn \theta_m \sn(\omega_b / 2) [} - \cs(\omega_m / 2) \cs(\omega_b / 2) \cs(\phi_m - \phi_b)] \\
g_{36} &= \sn(\omega_m / 2) \sn \theta_m [- \sn(\omega_m / 2) \cs \theta_m \sn \omega_b \sn(\phi_m - \phi_b) + \sn(\omega_m / 2) \sn \theta_m \cs \omega_b + \cs(\omega_m / 2) \sn \omega_b \cs(\phi_m - \phi_b)] / 2 \\
g_{44} &= 1 / 2 \\
g_{55} &= 1 - \cs \omega_b \\
g_{56} &= \sn^2(\omega_b / 2) \\
g_{66} &= 1 / 2
\end{align*}
The indices correspond to $\omega_m$, $\theta_m$, $\phi_m$, $\omega_b$, $\phi_b$, and $\omega_z$ in increasing order. Taking the square root of the determinant of the metric tensor shows that the differential volume element on the space is
\begin{equation*}
\diff \Omega = \frac{1}{16} \sin^2(\omega_m / 2) \abs{\sin \theta_m} \abs{\sin \omega_b} \diff \omega_m \diff \theta_m \diff \phi_m \diff \omega_b \diff \phi_b \diff \omega_z.
\end{equation*}
Integrating the volume element over the intervals $[0, 2 \pi]$, $[0, \pi]$, $[0, 2 \pi]$, $[0, 2 \pi]$, $[0, 2 \pi]$, and $[0, 2 \pi]$ shows that the volume of the space is $4 \pi^4$, as expected.

\begin{acknowledgments}
This is the acknowledgements.
\end{acknowledgments}

\bibliography{refs}

\end{document}